\documentclass[fleqn,usenatbib, openany]{mnras}

\usepackage[T1]{fontenc}
\usepackage{ae,aecompl}
\usepackage[utf8]{inputenc}
\usepackage{graphicx}	
\usepackage{rotating}
\usepackage{comment}
\usepackage{amsmath}	
\usepackage{amssymb}	
\usepackage{csquotes}
\usepackage{multirow}
\usepackage{verbatim}
\usepackage{booktabs}
\usepackage{graphics}
\usepackage{longtable}
\usepackage{soul}
\usepackage{newtxtext,newtxmath}
\usepackage{dblfloatfix}
\usepackage{float}
\usepackage[title]{appendix}

\newcommand{\etal}{et al.\ }

\title[Star-forming S0s in MaNGA]{Star-forming S0 Galaxies in SDSS-MaNGA: fading spirals or rejuvenated S0s?}

\author[Rathore \etal]
{Himansh Rathore,$^{1}$\thanks{E-mail: himansh@iitb.ac.in (HR)} Kavin Kumar,$^{2}$\thanks{E-mail: nrkavin18@iiserb.ac.in (KK)} Preetish K. Mishra,$^{3,4}$\thanks{Email: preetish@iucaa.in (PKM)} Yogesh Wadadekar,$^{3}$\thanks{E-mail: yogesh@ncra.tifr.res.in (YW)} and \newauthor Omkar Bait$^{3,5}$\thanks{Email: omkar.bait@unige.ch (OB)}
\\
$^{1}$Department of Physics, Indian Institute of Technology Bombay, Powai, Mumbai, Maharashtra
(400076), India \\
$^{2}$Department of Physics, Indian Institute of Science Education and Research Bhopal, Bhauri, Bhopal, Madhya Pradesh (462066), India \\
$^{3}$National Centre for Radio Astrophysics, Tata Institute of Fundamental Research, Post Bag 3, Ganeshkhind, Pune, Maharashtra (411007), India \\
$^{4}$ The Inter-University Centre for Astronomy and Astrophysics, Post bag 4, Ganeshkhind, Pune, Maharashtra (411007), India \\
$^{5}$ Observatoire de Gen\`eve, Universit\'e de Gen\`eve, 51 Ch. des Maillettes, 1290 Versoix, Switzerland \\}

\date{Accepted XXX. Received YYY; in original form ZZZ}

\pubyear{2021}

\begin{document}
\label{firstpage}
\pagerange{\pageref{firstpage}--\pageref{lastpage}}
\maketitle



\begin{center}
\begin{abstract}
We investigate the origin of rare star-formation in an otherwise red-and-dead population of S0 galaxies using spatially resolved spectroscopy. Our sample consists of $120$ low redshift ($z<0.1$) star-forming S0 (SF-S0) galaxies from the SDSS-IV MaNGA DR15. We have selected this sample after a visual inspection of deep images from the DESI Legacy 
Imaging Surveys DR9 and the Subaru/HSC-SSP survey PDR3, to remove contamination from spiral galaxies. We also construct two control samples of star-forming spirals (SF-Sps) and quenched S0s (Q-S0s) to explore their evolutionary link with the star-forming S0s. To study star-formation at resolved scales, we use dust-corrected $H_\alpha$ luminosity and stellar density ($\Sigma_\star$) maps to construct radial profiles of star-formation rate (SFR) surface density  ($\Sigma_{SFR}$) and specific SFR (sSFR). Examining these radial profiles, we find that star-formation in SF-S0s is centrally dominated as opposed to disc dominated star-formation in spirals. We also compared various global (size-mass relation, bulge-to-total luminosity ratio) and local (central stellar velocity dispersion) properties of SF-S0s to those of the control sample galaxies. We find that SF-S0s are structurally similar to the quenched S0s and are different from star-forming spirals. We infer that SF-S0s are unlikely to be fading spirals. Inspecting stellar and gas velocity maps, we find that more than $50\%$ of the SF-S0 sample shows signs of recent galaxy interactions such as kinematic misalignment, counter-rotation, and unsettled kinematics.  Based on these results, we conclude that in our sample of SF-S0s, star-formation has been rejuvenated, with minor mergers likely to be a major driver. 

\end{abstract} 
\end{center}

\begin{keywords}
galaxies: evolution,
galaxies: star formation,
galaxies: structure
\end{keywords}


\section{Introduction} \label{sec: introduction}

Several previous works and studies have shown that actively star-forming galaxies fall on a narrow sequence in the Star Formation Rate (SFR) vs stellar mass ($M_{\ast}$) parameter space. This is known as the Star Forming Main sequence (SFMS) and has been observed atleast upto redshits $z \sim 5$ (e.g. \cite{Brinchmann04, Daddi07, Salim07, Santini17}). A significant fraction of galaxies in the low-redshift Universe are known to deviate away from the SFMS towards lower star-formation rates and then evolve passively. There are multiple factors responsible for the termination of star-formation (also referred to as quenching) in galaxies, and the likelihood of a galaxy to be quenched seems to be a strong function of its morphology. Observations show that quenched galaxies tend to have early-type morphology \citep{Skibba2009, Bait17}. This does not come as a surprise as the early-types are typically massive galaxies living in dense environments \citep{Dressler1980, Goto2003}. They often harbour strong AGNs \citep{Schawinski2007, Kormendy&Ho2013} and have a  merger rich evolutionary history \citep{deLucia2006, Bournaud2007,Tapia2017, Eliche-Moral2018, Deeley2021}. Galaxy mergers, feedback from AGN and high density environment are identified as potent mechanisms for quenching of star-formation \citep{Croton2006, Gabor2010, Bluck2020}.

The picture of early-type galaxies being passive \lq red-and-dead\rq systems is now ingrained in our imaginations. However, studies in the past decade have uncovered a population of early-type galaxies exhibiting active star-formation. \cite{Schawinski09} found a sample of around $200$ blue early-type galaxies at $0.02 < z < 0.05$ exhibiting mild to moderate SFRs between $0.05$ and $50$ $M_\odot$ yr$^{-1}$. Even though these galaxies are extremely rare, constituting only about $\sim$5\% of the low redshift early-type galaxy population, their mere presence requires us to revise our understanding of the early-type population.

A number of studies have attempted to understand the origin of blue early-type galaxies. At $z \sim 0$, \cite{SK09} report blue early-types are increasingly abundant at low stellar masses, from being non-existent above $M_\ast \sim 10^{11} M_\odot$ to constituting $20-30\% $ of total early-type population below $M_\ast \sim 10^{9} M_\odot$. They infer that high mass ($M_\ast > 10^{10} M_\odot$) blue early-types are major merger remnants that are fading into the red-sequence. The low-mass ($M_\ast < 10^{10} M_\odot$) are probably building a star-forming disc. \cite{Wei09} have used H{\sc i} measurements for $27$ low-mass blue early-types and find that these galaxies have gas fractions similar to spirals in the blue cloud. They suggest that the star-formation in these galaxies is bursty, and involves externally triggered gas inflows possibly through minor mergers. Another study by \cite{Koshy17} using the \cite{Schawinski09} sample showed that the star-forming blue early-type galaxies at low redshifts could be normal ellipticals that may have undergone recent minor-merger events which supplied fresh gas to fuel star-formation.

It is clear from the above that the creation pathways of star-forming early-types are diverse and may have to do with the fact that they may not be a homogeneous class of objects. The early-type class includes spheroidal elliptical galaxies and discy S0 galaxies, both having different formation scenarios. The task of unravelling the mysteries of star-forming early-type population could become easier if one focuses on only one \lq type\rq of galaxies. In this work, we have chosen to study star-forming S0 galaxies.

The S0s are an interesting class of objects, traditionally placed in-between spirals and ellipticals on the Hubble tuning fork \citep{Hubble36}. They are now thought to be much closer to spirals than to ellipticals. The literature is full of observational works in support of S0s being transformed spirals \citep{Bergh1976,Salamanca2006, Barway2009,Laurikainen2010, Kormendy&Bender2012, Johnston14, Mishra2018, Rizzo2018} and there are also a number of proposed mechanisms to explain the transformation via gas starvation, galaxy mergers, tidal interaction and disc instability \citep{Bekki2002, Bekki2011, Querejeta2015, Saha2018}. However, it is important to note that the formation channels of S0s may depend on mass and environment of their progenitor galaxy \citep{McKelvie2018, Mishra19}. 

The structural and kinematic similarity of spirals and S0s evoke pointed questions regarding the origin of star-forming S0s. They might be a class of quenched S0 galaxies where the star-formation has been recently rejuvenated by some process, or they could be a population of star-forming spirals undergoing a last burst of star-formation before they turn into a typical S0 galaxy, or they could be a separate population with a distinct evolutionary history. 

In this work, we study star-forming S0 type galaxies at low redshift ($ z < 0.1$) using Integral Field Spectroscopy (IFS) data from the Sloan Digital Sky Survey's Mapping Nearby Galaxies at APO (SDSS - MaNGA) \citep{Bundy15, Blanton17} survey. The MaNGA survey provides us with spatially resolved spectroscopy of a galaxy thus enabling us to study star-formation in these galaxies in a spatially resolved fashion. This will help us uncover the site and spatial extent of star-formation in these objects. We also compare the resolved and global properties of star-forming S0s to a control sample of star-forming spirals and the typical quenched population of S0s. Such a comparison might connect this elusive class of objects to their progenitor population.  

The organisation of this paper is as follows. In Section \ref{sec: sample selection And Characterization}, we describe our sample selection and look at the three samples in more detail, and explore some of their global properties. In Section \ref{sec: data cube processing}, we describe the data-cube processing, and how resolved maps and radial profiles were constructed for the analyses of star-formation properties. We describe the main results related to the radial profiles in Section \ref{sec: results}. Further discussion on the whole study, and construction of an evolutionary history is carried out in Section \ref{sec: discussion}, and the main results are summarised in Section \ref{sec: conclusion}. Throughout, we adopt the Planck13 cosmology \citep{Planck13}, in which $H_0=67.8 \: \: {\rm km} \: {\rm s}^{-1} \: {\rm Mpc}^{-1}$, $\Omega_\Lambda = 0.69$ and $\Omega_m = 0.31$.

\section{Sample Selection and Characterisation} \label{sec: sample selection And Characterization}
In this section, we describe in detail the process of sample selection and its refinement, motivated by the problem at hand. We also describe some of the global properties of our sample galaxies.

\subsection{Sample Overview and Basic Selection Criteria} \label{sec:sample_overview}
We construct three galaxy samples:
\begin{itemize}
    \item Star-forming S0 (SF-S0) galaxies
    \item Star-forming spiral (SF-Sp) galaxies 
    \item Quenched S0 (Q-S0) galaxies
\end{itemize}

The latter two samples shall act as control samples, to which we shall compare our primary objects of interest - the star-forming S0s. The motivation for constructing the above control samples in addition to the sample of SF-S0s comes from the evolutionary possibilities presented in Section \ref{sec: introduction}. We wish to uncover the spatial distribution of star-formation in the SF-S0s as well as establish an evolutionary link (if present) between SF-S0s to either spirals or the general population of quenched S0s. 

All the galaxies across the three samples we construct were observed by the SDSS - MaNGA IFS survey \citep{Bundy15, Blanton17} and have been processed with the Pipe3D IFS data-processing pipeline \citep{SanchezI16, SanchezII16}. For SFR, stellar mass and redshift ($z$), we use the GSWLC-A2 (GALEX-SDSS-WISE LEGACY CATALOG) \citep{Salimcatalog16, SalimSBcatalog18} (hereafter referred to as the Salim catalogue). SFR and stellar mass are computed in the Salim catalogue by modelling the Ultra-Violet (UV), optical and mid Infra-Red (IR) broadband Spectral Energy Distribution (SED). We use only those objects which have a SED flag $=$ OK in the Salim catalogue. This criterion excludes galaxies that show broad emission lines, characteristic of broad-line AGNs. Since we are interested in studying nearby galaxies, we consider only those galaxies which have $z < 0.1$. 

For structural parameters - half light semi-major axis ($R_e$), axis ratio ($B/A$) and position angle (PA), we use the $r$-band PyMorph \citep{Vikram10} single S\'{e}rsic fits \citep{PM15}; hereafter PM15) for all galaxies belonging to SF-S0, SF-Sp and Q-S0 samples. Fits are performed in PM15 using SDSS DR15 images. 

For obtaining the morphological type of each galaxy we use the MaNGA DR15 deep-learning catalogue (hereafter DL15; \citep{PM15, DL15}, specifically the columns \lq TType\rq and \lq P\_S0\rq. The former is the morphological Hubble TType \citep{Nair10, deV63}, and the latter is the probability of an object being an S0 as opposed to being a pure elliptical. DL15 classifications of the MaNGA galaxies are based on SDSS DR15 images \citep{dr15datapaper}. Their deep-learning Convolutional Neural Networks (CNNs) are trained on two galaxy visual classification catalogues based on SDSS DR7 \citep{dr7datapaper} images - \cite{Nair10} and Galaxy Zoo - 2. DL15 claims a performance of $ > 90\%$ on the classifications provided by the deep-learning algorithm.  

Earlier studies (e.g., \cite{Schawinski09, Kaviraj09}) involving star-forming early type galaxies have used either optical or UV colours as a proxy for their star-formation activity. The optical colours are insensitive to low levels of star-formation activity \citep{Salim15}, and their use can result in wrong identification of star-forming galaxies as passive red objects \citep[e.g.,][]{Cortese2012}. The $NUV-r$ colour is shown to be good at capturing the ongoing star-formation in galaxies \citep{Salim15}. However, it does not take into account attenuation, which is known to vary with properties (stellar mass, SFR etc.) of galaxies \citep[e.g.,][]{Salim2020}. The last decade has provided us with large scale SED fits, with Salim catalogue being one of them. This allows us to select the SF-S0s, SF-Sps and Q-S0s directly by using a criteria based on the specific SFR (which is defined as the SFR per unit stellar mass) obtained from stellar population synthesis, instead of using colour as an (imperfect) proxy for star-formation. \cite{Salim15} suggests using
\begin{equation} \label{eq:sf_selection}
    \rm{sSFR} (\log(\rm{yr}^{-1})) \geq -10.8 
\end{equation}
for selecting star-forming objects, and
\begin{equation} \label{eq:q_selection}
    \rm{sSFR} (\log(\rm{yr}^{-1})) \leq -11.8 
\end{equation}
for selecting quenched objects. Objects with $-11.8 < \rm{sSFR} \: (\log(\rm{yr}^{-1})) < -10.8$ are said to reside in the green valley, which is the transition region between the SFMS and the quenched sequence \citep[see e.g.,][]{Bait17}.

We demand that a galaxy is considered for further analysis only if it has measurements of all the above mentioned quantities in the columns of the respective catalogues, and also has Pipe3D outputs. Hereafter, we shall refer to this as the basic selection criteria. This basic selection criteria does not introduce significant biases in our analysis, since the number of galaxies that are removed due to such cuts is very small (less than $\sim 5\%$). Moreover, we have also compared the distribution of stellar mass, luminosity, sSFR and morphology before and after applying these cuts. The distributions turn out to be very similar, indicating a low probability of significant bias due to the use of the basic selection criteria.

In order to maintain some uniformity across all samples we restrict the stellar mass range of SF-Sps and Q-S0s to be the same as that of SF-S0s, since stellar mass is a fundamental parameter in driving galaxy evolution \citep{Peng2010}. But as we shall realise later, this will turn out to be a feeble constraint given that we find a large stellar mass range is spanned by the SF-S0s.

\subsection{Sample Of Star-forming S0s (SF-S0s)} \label{sec: SFS0_selection}
We select those galaxies that pass the basic selection criteria and have TType $\leq 0$ and P\_S0 $> 0.5$ as recommended by DL15 for identifying S0s. We select the star-forming objects out of these by using the \cite{Salim15} criterion (equation (\ref{eq:sf_selection})), yielding $211$ galaxies at this stage. 

Faint spiral structure can be missed in shallow imaging, and thus a spiral galaxy can be mis-classified as S0. We do a strict visual inspection of colour composite images deeper than the SDSS for each of these objects in order to remove spiral contaminants from our sample. We first check the very deep Hyper Suprime-Cam Subaru Strategic Program (HSC-SSP) PDR3 \citep{Aihara2021} images from the hscMap\footnote{\url{https://hsc-release.mtk.nao.ac.jp/hscMap-pdr3/app/}}. We find that only $15$ out of the $211$ objects have HSC PDR3 images. We then look at the deep images from the DESI Legacy Survey DR9 \citep{Schlegel2021} from the sky viewer\footnote{\url{https://www.legacysurvey.org/viewer/}}, which were available for almost all of the $211$ objects. For galaxies where DESI Legacy DR9 images were not available, we obtained image cut-outs from the PanSTARSS-1 DR2\footnote{\url{https://ps1images.stsci.edu/cgi-bin/ps1cutouts}} \citep{Panstarrs18}. In addition to removing galaxies which showed signatures of spiral structure we removed those which showed visually apparent signatures of ongoing merger, since merging systems have a disturbed morphology. We also discarded visually edge-on objects during this visual inspection, since it is difficult to identify whether they are spirals or S0. An edge-on spiral galaxy with a large bulge
can easily be mis-classified as S0, because at sufficiently oblique angles spiral structure is often invisible.

Lacking the bright blue star clusters that illuminate the discs of spirals, the discs of S0s tend to be smooth, making it hard to distinguish S0s from ellipticals, especially in case of face-on S0s lacking other discy features such as bars, rings etc. Some genuine S0s may have dropped out of our sample due to this effect. 

During this visual inspection, the first two authours independently classified each galaxy as S0 or spiral based on the aforementioned procedure. A discussion amongst them took place for conflicting classifications ($\sim10\%$ of the cases) and a common consensus was reached. Other authors were consulted if disagreement still remained. After this, other authors also inspected the sample finally shortlisted by the first two authors, and no conflicts were found.

On the completion of the above visual inspection, we are reduced to $120$ galaxies, comprising our sample of SF-S0s. Around $30\%$ of the galaxies that got classified as S0 based on DL15 classifications from SDSS imaging showed signatures of spiral structure in deeper images. This reinforces the need to perform a strict visual inspection for studies involving S0 galaxies that utilise shallow imaging, so that a clean sample can be obtained. Our star-forming S0 sample is expected to be curated and reliable, in this respect.
\begin{figure}
    \centering
    \includegraphics[width = 0.15\textwidth]{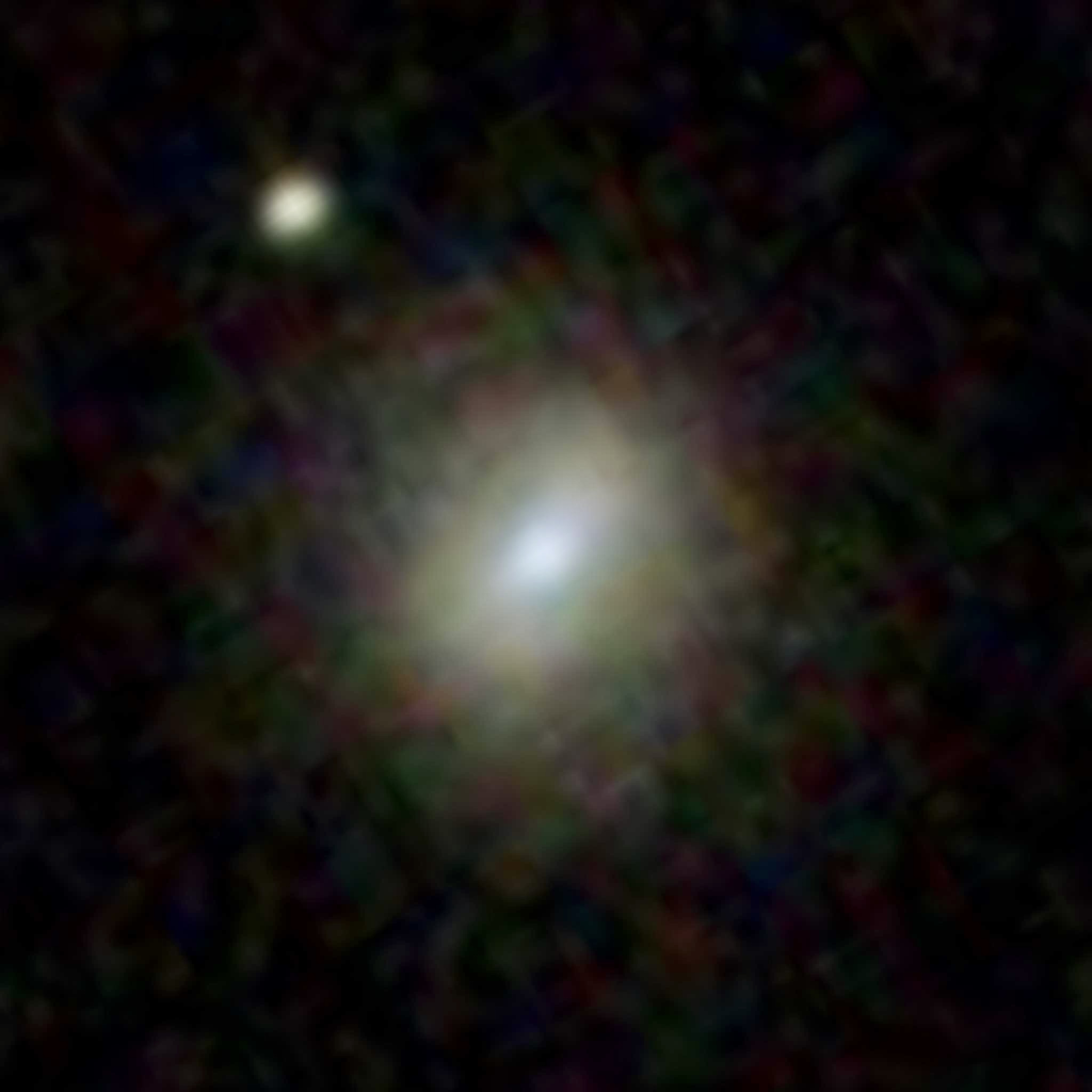}
    \includegraphics[width = 0.15\textwidth]{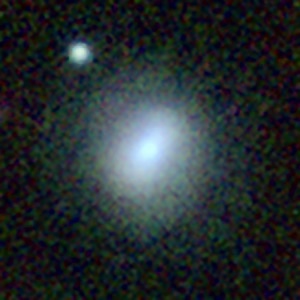}
    \includegraphics[width = 0.15\textwidth]{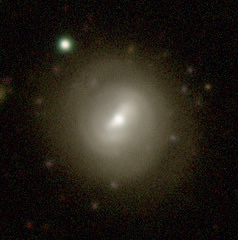}
    \caption{Left to right: Comparison of images of a galaxy (PLATEIFU 8552-3701) in SDSS DR14, DESI Legacy DR9 and HSC PDR3 surveys. As the depth of the image improves, spiral features start becoming prominent in galaxies that were otherwise classified as S0 based on SDSS imaging. Size of each image is $40$ arcsec $\times$ 40 arcsec.}
    \label{fig: img_comparison}
\end{figure}

Fig. \ref{fig: img_comparison} shows an example of a galaxy that would be classified as S0 based on SDSS DR14 imaging, but spiral structure starts becoming evident in the deeper DESI Legacy DR9 images, and becomes even more prominent in the even deeper HSC PDR3 images. The Legacy DR9 and HSC PDR3 image cutouts for the SF-S0 galaxies are presented in Appendix A. Two examples of catastrophic failure of the DL15 classification algorithm are also shown in Appendix A. Based on SDSS DR14 imaging, these two galaxies are clearly spirals, but have been mis-classified as S0. Thus, for studies that utilise deep-learning morphological classifications, we recommend researchers to visually inspect the images, if feasible. This is specially important for studies of rare objects where sample size is relatively small and it is imperative to have a clean sample.

The stellar mass and redshift range of SF-S0s is $8.75 < \log \frac{M_\ast}{M_{\odot}} < 11.25$ and $0.01 < z < 0.08$ respectively.

\subsection{Control Sample Of Quenched S0s (Q-S0s)}
 We select S0s according to the DL15 criteria and use the \cite{Salim15} criteria for the selection of quenched objects (equation (\ref{eq:q_selection})). We apply the basic selection criteria, and carry out a visual inspection of the $254$ galaxies obtained so far using the same methodology as was used for SF-S0s. This leaves us with $234$ objects. We found very few spiral contaminants, which is consistent with the fact that truly passive spirals are very rare \citep{Cortese2012}. We impose the stellar mass range of this sample to be the same as that of SF-S0s, which slightly reduces the sample size to $227$. These $227$ objects shall comprise the control sample of quenched S0s (Q-S0s). 
 
\subsection{Control Sample Of Star-forming Spirals (SF-Sps)}
We select late-type galaxies using the DL15 criteria TType $> 0$ and select star-forming objects using the \cite{Salim15} criteria (equation (\ref{eq:sf_selection})). After this we apply the basic selection criteria, and are left with $1508$ galaxies. $8$ out of these $1508$ galaxies are designated an unusual TTYPE $= 10$ in DL15. Upon inspecting these $8$ galaxies, we realised that they are mostly disturbed, with some of them possessing an irregular morphology. We have removed these objects, and are thus left with $1500$ galaxies. We impose the stellar mass range of this sample to be the same as that of SF-S0s, which slightly reduces the sample size to $1468$. These $1468$ objects shall comprise the control sample of star-forming spirals (SF-Sps). We have not carried out a visual inspection of the images of SF-Sps, given the large sample size. Moreover, it is difficult to mis-classify any other morphology as a spiral. Spiral features that are detected in shallow imaging will become more prominent in deeper images, but the galaxy will obviously still be said to have a spiral morphology. Given the accuracy of the DL15 algorithm, we expect less than $10\%$ mis-classifications in our SF-Sp control sample. At this stage, some edge-on spiral galaxies might contaminate our SF-Sp sample due to the same reason mentioned in Section \ref{sec: SFS0_selection}. But during the analysis of resolved properties, such galaxies will automatically be rejected through a formal inclination cut to be applied later (Section \ref{sec: radial profiles}).
\\
\\
Next, we describe some global properties that characterise our main sample of SF-S0s, and the control samples of SF-Sps and Q-S0s.

\subsection{Global Properties} \label{sec: global props}
\begin{figure}
    \centering
    \includegraphics[width = 0.45\textwidth]{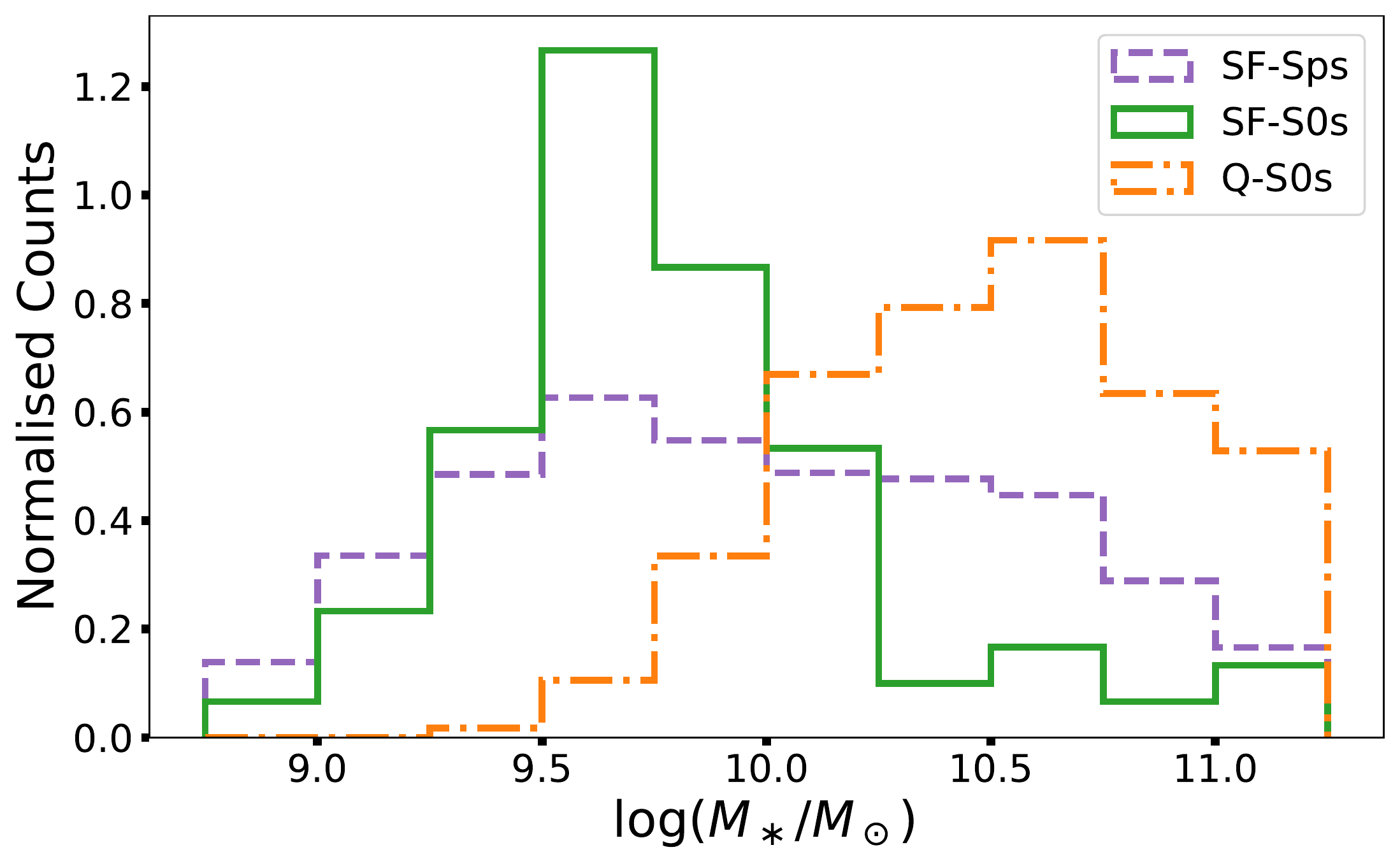}
    \caption{The stellar mass ($\log \frac{M_\ast}{M_{\odot}}$) histogram for the three samples. The control sample of star-forming spirals (SF-Sps) having $1468$ objects is depicted by the purple dashed line, the sample of star-forming S0s (SF-S0s) having $120$ objects is depicted by the green solid line, and the control sample of quenched S0s (Q-S0s) having $227$ objects is depicted by the orange dash-dot line. The stellar mass range of SF-Sps and Q-S0s has been imposed to be the same as that of SF-S0s. On average, Q-S0s have a higher stellar mass as expected of quenched early type galaxies. Most of the SF-S0 galaxies have $\log \frac{M_\ast}{M_{\odot}}< 10$, consistent with the findings of previous studies. Area under all histograms is normalised to unity.}
    \label{fig: sm_hist_combined}
\end{figure}

Fig. \ref{fig: sm_hist_combined} shows the stellar mass histogram for SF-Sps, SF-S0s and Q-S0s. The median $\log \frac{M_\ast}{M_{\odot}}$ for the three samples are $9.93$, $9.72$ and $10.54$ respectively. Q-S0s having a higher stellar mass than SF-Sps on average is consistent with the fact that quenched early type galaxies tend to be more massive (e.g., \cite{Baldry04a, Con06, Buit13, Bait17}). Most SF-S0s have $\log \frac{M_\ast}{M_{\odot}} < 10.25$  , which is consistent with findings of previous studies involving blue early type galaxies (e.g., \cite{SK09, Huertas10}). 

We note that we find very few Q-S0s at lower stellar masses. In addition to the fact that quenched galaxies in general tend to have higher stellar masses, we believe selection effects are also present in our sample, which significantly reduces the number of quenched objects. Quenched galaxies at low stellar masses will have very low luminosities, particularly in the bluer bands, and thus very low flux at a given redshift. This will make it very difficult to obtain stellar population estimates for such galaxies from UV-Optical-mid IR SED fits. Hence, such galaxies might have been automatically excluded from our sample. 

The MaNGA survey also has complicated selection effects \citep{Bundy15} driven both by technological constraints such as the number of deployable IFUs of different sizes, as well as science driven constraints such as the requirement to have a flat number density distribution in absolute $i$-band magnitude. Fortunately, correction factors in the form of volume weights have already been tabulated for MaNGA galaxies \citep{Wake17}. These multiplicative weights need to be applied when any statistical quantity is being calculated for a set of MaNGA galaxies. We have applied these weights in our analysis to make our results statistically more rigorous. From this point onwards, whenever we refer to a statistical quantity as being weighted, the weights used correspond to these volume correction factors\footnote{A few galaxies (less than $5\%$ of a particular sample) will get dropped out because they do not possess volume weights.}.

\begin{figure}
    \centering
    \includegraphics[width = 0.45\textwidth]{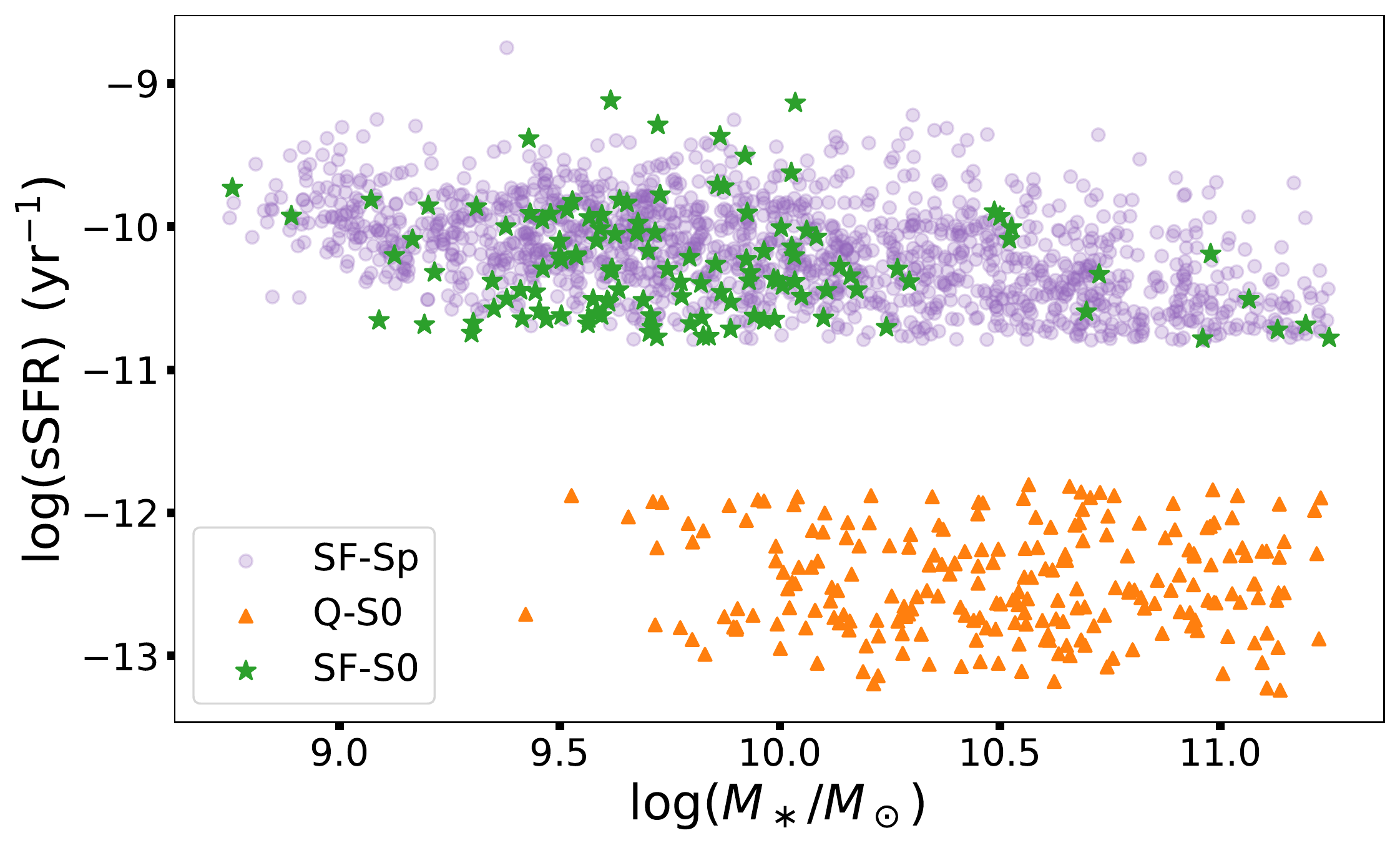}
    \caption{The Star Formation Main Sequence (SFMS) plot for the three samples. The control sample of star-forming spirals (SF-Sps) having $1468$ objects is depicted by purple circles, the sample of star-forming S0s (SF-S0s) having $120$ objects is depicted by green stars, and the control sample of quenched S0s (Q-S0s) having $227$ objects is depicted by orange triangles. On the basis of median sSFR values, SF-S0s are slightly less star-forming on average as compared to SF-Sps.}
    \label{fig: sfr_sm_combined}
\end{figure}

Fig. \ref{fig: sfr_sm_combined} shows the Star Forming Main Sequence (SFMS). SF-Sps and SF-S0s seem to have similar sSFR distributions, with median values in units of $\log(yr^{-1})$ being $-10.16$ and $-10.32$ respectively. SF-S0s are thus somewhat less star-forming on average as compared to SF-Sps.

The stellar mass histogram for SF-S0s (Fig. \ref{fig: sm_hist_combined}) suggests a sharp decline in their population at stellar masses higher than $\log \frac{M_\ast}{M_{\odot}}\sim 10.25$. Thus, we split our three samples at $\log \frac{M_\ast}{M_{\odot}}= 10.25$ and study objects with $\log \frac{M_\ast}{M_{\odot}}\leq 10.25$ (low stellar mass) and $\log \frac{M_\ast}{M_{\odot}}> 10.25$ (high stellar mass) separately. The galaxy counts in each mass bin for the three samples are shown in Table \ref{tab:num_stats}.
\begin{table}
    \centering
    \caption{Number counts for the sample of star-forming S0s (SF-S0s), star-forming spirals (SF-Sps) and quenched S0s (Q-S0s).}
    \label{tab:num_stats}
    \begin{tabular}{|c|c|c|}
    \hline
    Sample & $\log(M_\ast/M_\odot) \leq 10.25$ & $\log(M_\ast/M_\odot) > 10.25$\\
    \hline
    SF-S0 & $106$ & $14$\\
    SF-Sp & $964$ & $504$\\
    Q-S0 & $64$ & $163$\\
    \hline
    \end{tabular}
\end{table} 

In the next section, we describe post-processing of the Pipe3D data-cubes  which will provide us with radial profiles of star-formation, and thus help us determine where the star-formation is happening in these SF-S0s.  

\section{Data Cube Processing} \label{sec: data cube processing}

IFS observations of MaNGA galaxies have been reduced by Pipe3D, which is a fitting pipeline for determining stellar population and ionised gas properties for IFS data. A full description of the pipeline can be found in \cite{SanchezI16, SanchezII16}. 

In brief, a spatial binning of the spaxels (short for spatial pixels) in the MaNGA data-cubes is performed to obtain a high S/N of the stellar continuum. Then, stellar population and strong emission lines are analysed quasi-simultaneously. First, a template of Simple Stellar Populations (SSPs) are fitted to each spatial bin, giving the kinematics and dust attenuation of the stellar population. This fitted continuum is then subtracted from the original spectrum to yield an emission line only data-cube. The strong emission lines are fitted with Gaussian profiles, and this model of emission lines is then subtracted from the original spectrum. This removes the effect of strong emission lines, and then this subtracted spectrum is fitted using a stellar library to derive the stellar population properties like age, metallicity and star-formation history. A Monte-Carlo approach is used to derive the stellar population parameters, thus yielding error estimates as well. Weak emission lines are analysed separately using a different procedure, which involves direct estimation of their parameters without Gaussian fitting. Spatial de-zonification is then performed to decouple the analysis of emission lines and stellar continuum, and also to recover stellar population and emission line properties at the spaxel level. 

In addition to MaNGA, Pipe3D has been used for processing data from other IFS surveys like CALIFA and SAMI \citep{SanchezI16, SanchezII16, CALIFA12, Croom21, Allen15}, and several previous works utilising the MaNGA survey have used Pipe3D products to carry out their analyses (e.g., \cite{Ellison18,Bluck2020}).  

In this work, we use the publicly available Pipe3D\footnote{\url{https://www.sdss.org/dr16/manga/manga-data/manga-pipe3d-value-added-catalog/}} value added catalog, in particular, the resolved maps of stellar population and emission line properties. These data-cubes are 3-dimensional arrays, whose two axes represent the spaxel coordinates and each layer along the third axis contains the resolved maps of various parameters. In this section, we describe processing of these data-cubes in order to obtain resolved maps and radial profiles of the quantities of interest. 

\subsection{Resolved Maps} \label{sec: resolved maps}

We study surface density of SFR ($\Sigma_{SFR}$) on resolved scales instead of the SFR, since spaxels for galaxies at different redshifts will probe different physical areas. Hence, it is required to normalise the SFR with the physical area. This approach has been used in several previous studies (e.g., \cite{Ellison18, Wang19, Bluck2020}). Following \cite{Ellison18}, we compute the SFR for each spaxel from dust-corrected $H_\alpha$ luminosity maps, by using the \cite{Kennicutt98} calibration with a Salpeter stellar Initial Mass Function (IMF) \citep{Salpeter55}. The same IMF has been used in the Pipe3D Fits \citep{SanchezII16}. 

Pipe3D provides the observed $H_\alpha$ flux maps, which need to be corrected for dust extinction. Following \cite{Ellison18, Wang19}, we use the method of Balmer decrements to estimate the nebular colour excess, by assuming an intrinsic Balmer decrement $\left(H_\alpha/H_\beta \right)_{int} = 2.86$, corresponding to Case B recombination with a temperature of $10^4$ K and electron density $n_e = 10^2$ cm$^{-3}$ \citep{Osterbrock1989}. Following \cite{Dominguez13}, we obtain the attenuation of $H_\alpha$ from the nebular colour excess by using the \cite{Calzetti00} reddening curve.

We convert the dust corrected $H_\alpha$ flux of each spaxel to $H_\alpha$ luminosity using the luminosity distance corresponding to our chosen cosmology. We then convert the $H_\alpha$ luminosity of each spaxel to its SFR using the \citet{Kennicutt98} calibration. Using the angular size of MaNGA spaxels ($0.5$ arcsec) and the galaxy redshift, we determine the physical area of a spaxel (in kpc$^2$) for each galaxy by using our chosen cosmology. $\Sigma_{SFR}$ of a spaxel in units of ($M_\odot$ yr$^{-1}$ \rm{kpc}$^{-2}$) is simply given by the ratio of SFR of a spaxel and its area.

We calculate the error in $\Sigma_{SFR}$ using the $H_\alpha$ flux error maps provided by Pipe3D, propagated to the \citet{Kennicutt98} calibration. We do not apply any dust-correction for these. Pipe3D provides dust corrected stellar mass surface density ($\Sigma_\ast$) maps (in units of $M_\odot$ kpc$^{-2}$), and we use them directly.

During the selection of our three samples (Section \ref{sec:sample_overview}), we have considered only those galaxies that have the Salim catalogue SED fit flag $=$ OK. This automatically excludes galaxies that have broad emission lines characteristic of broad-line AGNs. But still, AGNs with narrow emission lines might remain, and can contaminate the resolved analysis because some spaxels will then have AGN activity as the dominant emission line source instead of star-formation. 

The Baldwin, Phillips \& Telervich diagram (BPT) \citep{BPT81} is commonly used to identify whether the dominant source of emission lines in a galaxy spectrum is due to star-formation or due to an AGN. Such an analysis can also be done on resolved scales (e.g., \cite{Wang19, Bluck2020}) and we can identify the dominant source of emission lines for each spaxel. We classify each spaxel as either AGN (Seyfert or LINER), or SF (Star Forming), or Composite, depending on its position in the BPT parameter space \citep{Kauff03, Kew06}. 

For any resolved analysis involving star-formation rates, we consider only the \lq star-forming spaxels\rq, which we henceforth define as spaxels which are categorised as SF in the resolved BPT diagram, and have Signal to Noise Ratio (S/N) $\geq 3$ in the $4$ emission lines we have used for the BPT diagnostic ($H_\alpha \: \lambda6563$, $H_\beta \: \lambda4861$, [NII] $\lambda6584$, [OIII] $\lambda5007$). 

We have eye-balled the resolved BPT diagrams of the SF-S0 galaxies, and realised $\sim 77\%$ of the galaxies were dominated by SF spaxels, $15\%$ were dominated by Composite spaxels, and only $\sim 8\%$ were dominated by AGN (Seyfert or LINER) spaxels. Thus, most of the SF-S0 sample has star-formation activity as the dominant source of emission lines. 

Next, we describe our procedure to obtain radial profiles of the quantities of interest.

\subsection{Radial Profiles} \label{sec: radial profiles}
Many recent works (e.g., \cite{Bluck2020, Santucci20, Wang19, Ellison18}) have demonstrated the importance of quantities such as SFR surface density and specific SFR to study galaxy evolution at resolved scales using IFS data. A convenient way to study the variation of physical quantities of interest across the galaxies is through radial profiles. These are essentially a plot of azimuthally averaged-out quantities as a function of distance from the centre of a galaxy. Below, we describe our general method for obtaining radial profiles for our sample galaxies.

We construct elliptical apertures over the map of the desired quantity which are centered at the centre of the galaxy \footnote{We have taken centre of the IFS field of view instead of centre of the galaxy, under the assumption that they coincide for almost all of the MaNGA sample.} and set the ellipticity and position angle of these apertures (taken from PM15\footnote{In PM15, ellipticity has been defined as $1 - \frac{b}{a}$, where $\frac{b}{a}$ is the semi-minor to semi-major axis ratio.}) to be the same as that of the galaxy. The central elliptical aperture was chosen to have a semi-major axis of $1$ kpc. We then kept making subsequent annuli having a thickness of $1$ kpc along the direction of the major axis till the semi-major axis of outermost ellipse reached the boundary of the IFS Field of View of that galaxy\footnote{The size of IFS Field of View of a galaxy can be inferred from its PLATEIFU \citep{Bundy15}}.

\begin{figure}
    \centering
    \includegraphics[width = 0.45\textwidth]{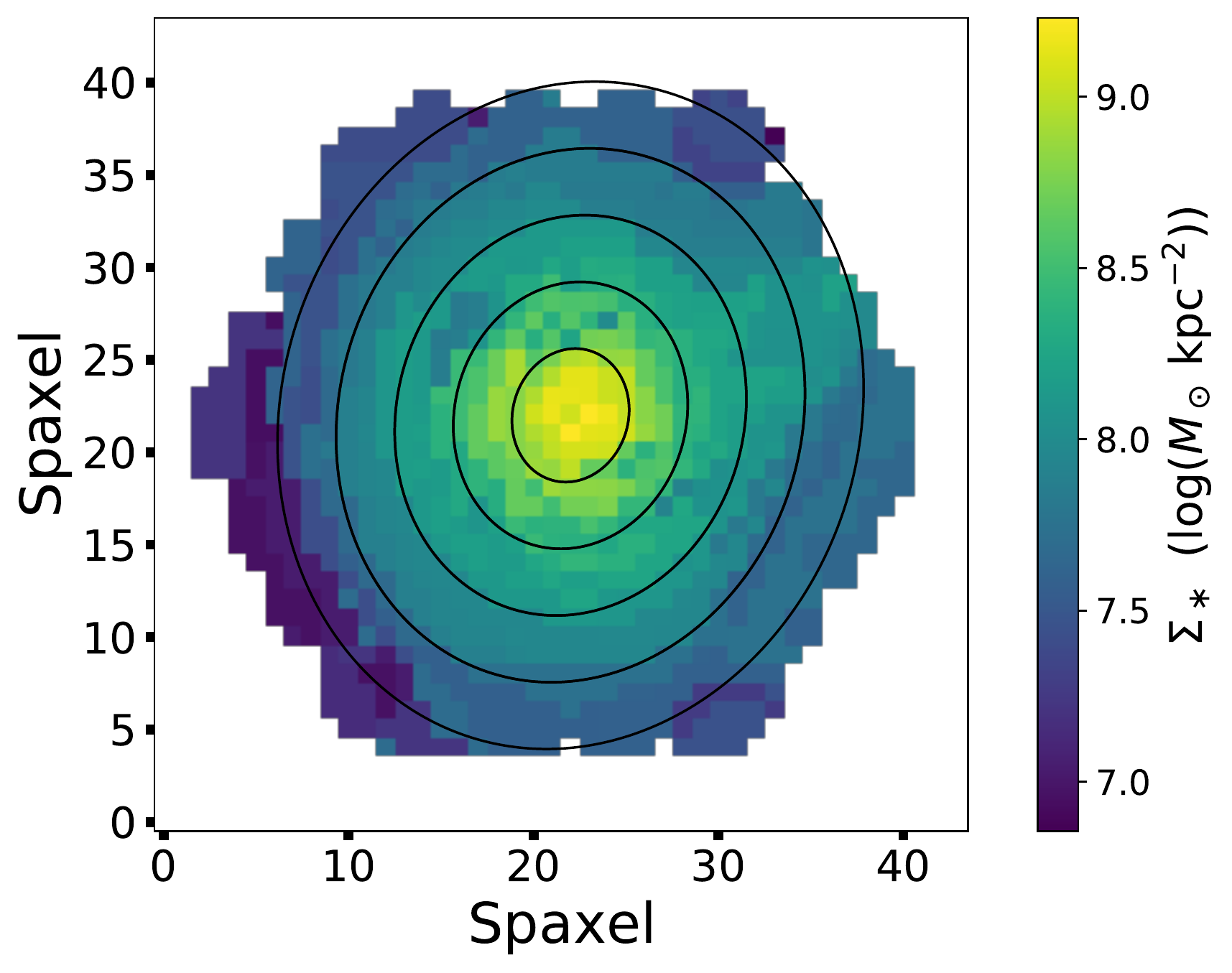}
    \caption{Apertures drawn over the stellar mass surface density ($\Sigma_\ast$) map of a SF-S0 (star-forming S0) galaxy with PLATEIFU 8249-3703. The ellipticity and position angle of the apertures is same as that of the galaxy. The central aperture has a semi-major axis of $1$ kpc and subsequent annuli have a thickness of $1$ kpc along the direction of major axis (see Section \ref{sec: radial profiles} for more details). Colour depicts the $\Sigma_\ast$ value for each spaxel.}
    \label{fig: aper_eg}
\end{figure}

\begin{figure}
    \centering
    \includegraphics[width = 0.40\textwidth]{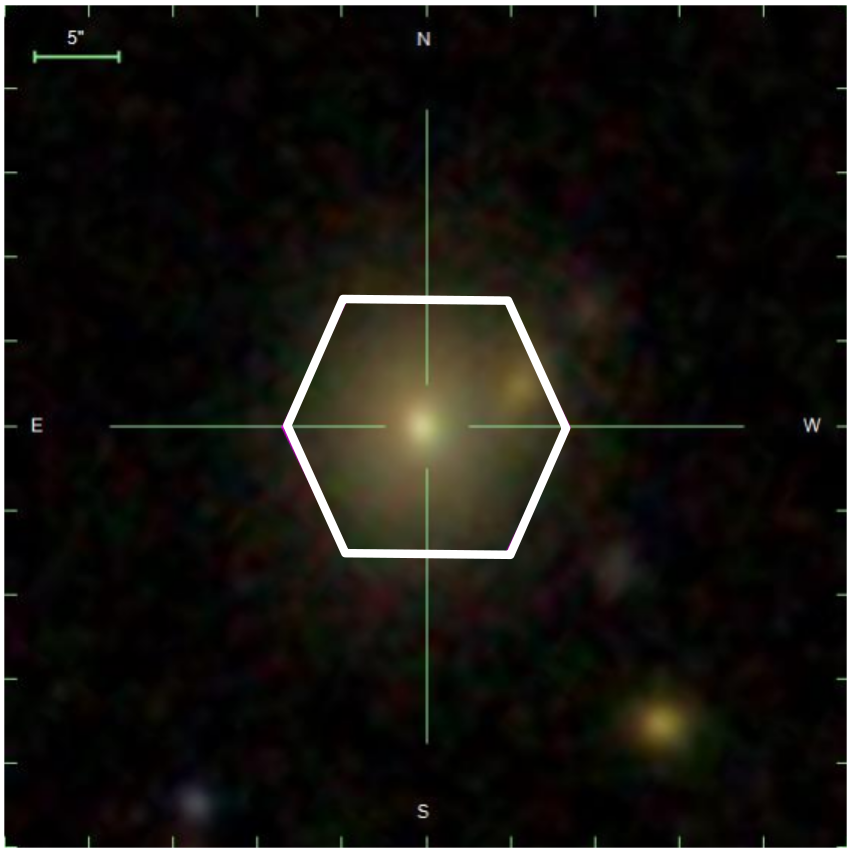}
    \caption{The image of the galaxy with PLATEIFU 8249-3703 along with the hexagonal MaNGA Integral Field Unit. Scale shown on top left is $5$ arcsec.}
    \label{fig:IFU_img}
\end{figure}

We use the Astropy affiliated package Photutils (photutils.aperture) \citep{photutils1.0.1,astropy:2013, astropy:2018} to construct the apertures and compute the radial profile of the desired quantity. An example is shown in Fig. \ref{fig: aper_eg}, wherein apertures are constructed over the $\Sigma_\ast$ map of a SF-S0 galaxy with PLATEIFU 8249-3703. The image of this galaxy along with the hexagonal MaNGA Integral Field Unit, which is obtained from the SDSS-MaNGA web interface MARVIN \footnote{\url{https://dr15.sdss.org/marvin/}} \citep{Marvin} is shown in Fig. \ref{fig:IFU_img}.

Quantities that are surface densities will be affected by inclination of the galaxy, hence de-projection is required. We derive inclination angles using the axis ratio \citep{Willick99} as shown in equation (\ref{eq:inclination}) below: 
\begin{equation} \label{eq:inclination}
    \cos^2{i} = \left\{
\begin{array}{ll}
      \frac{(1 - \epsilon)^2 - (1 - \epsilon_{max})^2}{1 - (1 - \epsilon_{max})^2} , & \epsilon < \epsilon_{max} \\
      0 , & \epsilon > \epsilon_{max} \\
\end{array} 
\right.
\end{equation}
, where $i$ is the inclination angle, $\epsilon$ is the ellipticity of the galaxy. $\epsilon_{max}$ is the approximate ellipticity exhibited by an edge-on spiral, which is taken to be $0.8$ \citep{Willick99}.

We scale (multiply) the value of each spaxel by the change in surface area probed by a spaxel as compared to the case when the galaxy was face-on. This scaling factor turns out to be $\sec{(i)}$, which can be derived from simple trigonometry. Note that the key assumption in this formalism is that the outermost isophote of disc galaxies is intrinsically circular, and inclination changes only the lengths that are perpendicular to the direction of the major axis. For a study of radial profiles, we consider only those galaxies that have inclination less than $75^\circ$, so that the de-projection formulation is reliable. That means the intrinsic thickness of the disc should not play a significant role. As a result of this inclination cut, we are left with $117$ objects in the SF-S0 sample (out of which $104$ have $\log(M_\ast/M_\odot) \leq 10.25$) and $1171$ objects in the SF-Sp sample (out of which $751$ have $\log(M_\ast/M_\odot) \leq 10.25$). We do not construct radial $\Sigma_{SFR}$ and sSFR profiles for the Q-S0s, since they are quenched galaxies and thus very few spaxels will have detection of emission lines above the required S/N threshold.

In the next section, we present the radial profiles of $\Sigma_{SFR}$ and sSFR for SF-S0s and SF-Sps, with the goal of determining the location and extent of star-formation in the SF-S0s.

\section{Results} \label{sec: results}
In this section, we present the radial profiles of $\Sigma_{SFR}$ and sSFR for SF-S0s and SF-Sps. We generate them as follows:
\begin{itemize}
    \item First, we construct apertures over $\Sigma_{SFR}$/sSFR map of each galaxy and take the mean over spaxels lying in a radial bin. This gives the radial profile of a single galaxy
    \item Then, we compute the weighted median radial profile for SF-S0 and SF-Sp samples
    \item For each galaxy, we compute the error of $\Sigma_{SFR}/sSFR$ in a radial bin by taking the error on the mean. The typical error for the radial profile of a given sample (SF-S0s/SF-Sps) is calculated by taking the weighted median across the errors on individual radial profiles
\end{itemize}
We obtain the sSFR map of a galaxy by taking ratio of $\Sigma_{SFR}$ and $\Sigma_\ast$ maps spaxel-by-spaxel.

For objects with $\log{M_\ast/M_\odot} \leq 10.25$, we construct these radial profiles out to a radial distance of $3.5$ kpc only, since beyond that there are not enough high S/N spaxels of SF-S0s. For objects with $\log{M_\ast/M_\odot} > 10.25$, we restrict the radial profiles upto $\mathbf{7.5}$ kpc for the same reason.

\subsection{SFR Surface Density Radial Profiles} \label{sec: sigma_sfr radial profile}

\begin{figure*}
    \centering
    \includegraphics[width = 1\textwidth]{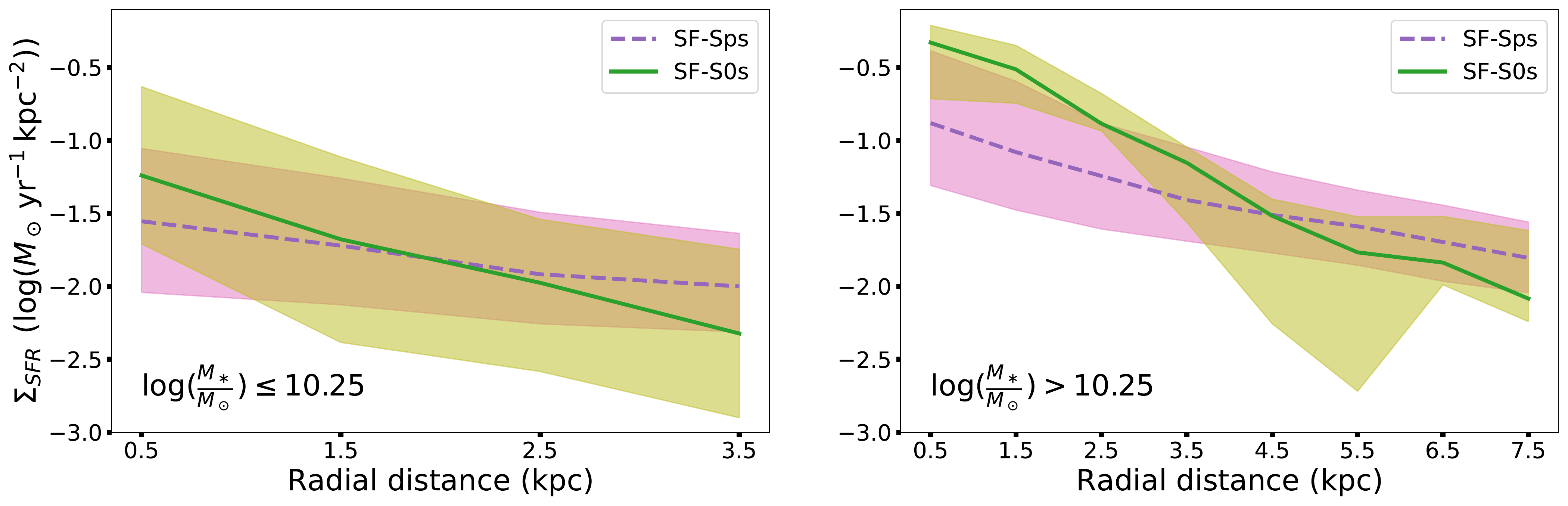}
    \caption{$\Sigma_{SFR}$ radial profile for star-forming S0s (SF-S0s, green solid line) and star-forming spirals (SF-Sps, purple dashed line) for both low stellar mass (left panel) and high stellar mass (right panel) bins. $25 - 75$ percentile bands depicting the scatter of a particular sample are also shown}. The typical measurement error on the radial profiles is very small ($\sim 0.001$ dex). $\Sigma_{SFR}$ radial profile for SF-S0s is centrally peaked, and declines relatively more rapidly than that of SF-Sps.
    \label{fig: sigma_sfr_profile_with_typical_errors}
\end{figure*}

Figure \ref{fig: sigma_sfr_profile_with_typical_errors} shows the $\Sigma_{SFR}$ radial profile for SF-Sps and SF-S0s, for both stellar mass bins. The coloured lines denote the weighted median $\Sigma_{SFR}$ as a function of radial distance from the galaxy centre and the coloured bands enclose $25-75$ weighted percentile of the profiles. The typical error on $\log(\Sigma_{SFR})$ is negligible (0.001 dex), since we have only used high S/N $H_\alpha$ spaxels.

We observe that $\Sigma_{SFR}$ is centrally peaked in SF-S0s and declines with radial distance. The decline seems relatively more rapid for SF-S0s than for SF-Sps. For the low stellar mass bin, the central $\Sigma_{SFR}$ of SF-S0s is comparable or slightly greater than that of SF-Sps. For the high stellar mass bin, the central $\Sigma_{SFR}$ of SF-S0s is definitely greater than that of SF-Sps. Low stellar mass SF-S0s have relatively lower $\Sigma_{SFR}$ values in the outer regions, as compared to SF-Sps.

\subsection{sSFR Radial Profiles} \label{sec: ssfr radial profiles}

\begin{figure*}
    \centering
    \includegraphics[width = 1\textwidth]{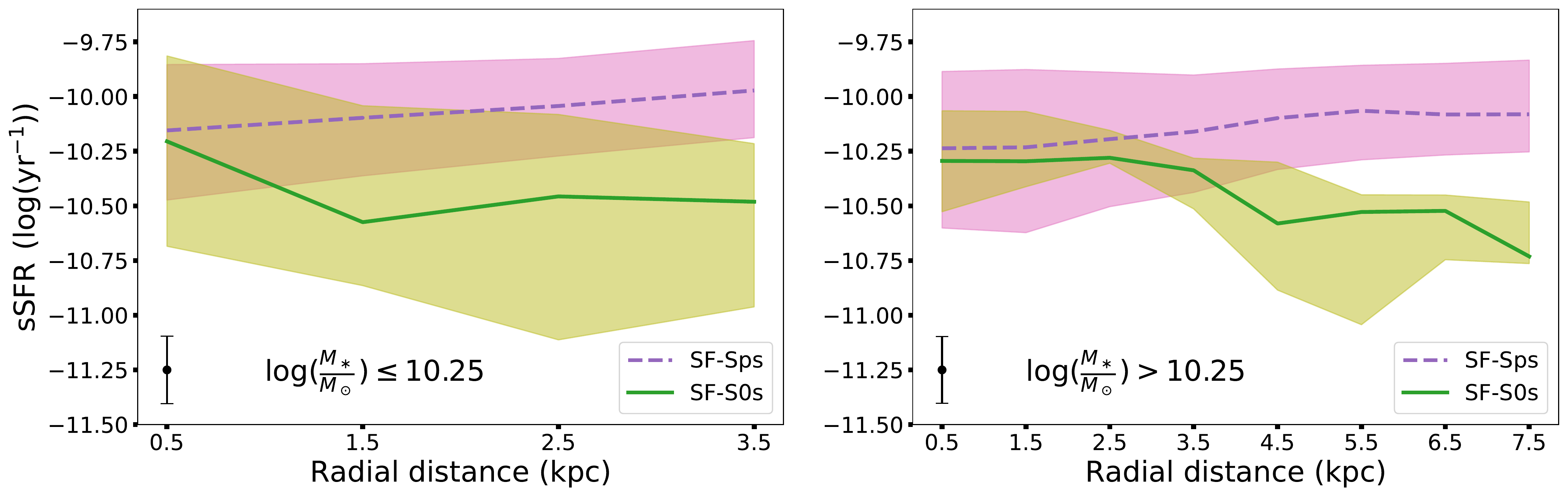}
    \caption{sSFR radial profile for star-forming S0s (SF-S0s, green solid line) and star-forming spirals (SF-Sps, purple dashed line) for both low stellar mass (left panel) and high stellar mass (right panel) bins. The typical error bar on the profile is shown in black in the bottom left corner. $25 - 75$ percentile bands depicting the scatter of the population are shown for each sample.} sSFR radial profile of SF-S0s is centrally peaked, and has a negative overall slope. Whereas, the sSFR profile of SF-Sps has a positive overall slope.
    \label{fig: ssfr_profile_with_typical_errors}
\end{figure*}

Figure \ref{fig: ssfr_profile_with_typical_errors} shows the sSFR radial profile for SF-S0s and SF-Sps for both stellar mass bins. The coloured lines denote the weighted median sSFR as a function of radial distance from the galaxy centre and the coloured bands enclose $25-75$ weighted percentile of the profiles. The typical error (bottom left corner) is the combination of uncertainties on $\Sigma_{SFR}$ and $\Sigma_\ast$. The uncertainty on $\log(\Sigma_\ast)$ is $\sim 0.15$ \citep{SanchezI16, SanchezII16}, almost all of it being systematic (Sebastian S{\'a}nchez, private communication).

sSFR for the SF-S0s is highest in the centre and declines on average with the radial distance. This again is suggestive of SF-S0s possessing centrally dominated star-formation. In the central regions, SF-S0s and SF-Sps have similar sSFR values. Whereas, in the outer regions sSFR of SF-S0s is lower than that of SF-Sps. For the SF-Sps, we observe a rising sSFR radial profile, which is consistent with previous IFS studies \citep{GonzalezDelgado16}. This suggests that dominant star-formation in spiral galaxies happens in the disc.
\\
\\
From the $\Sigma_{SFR}$ and sSFR radial profiles, it is evident that star-formation in SF-S0s is primarily happening in the central regions, as opposed to disc dominated star-formation in SF-Sps. We caution that it is difficult to give strong statistical conclusions for high stellar mass SF-S0s, due to small number of objects. In the next section (Section \ref{sec: discussion}), we attempt to build an understanding of the evolutionary history of these rare, star-forming S0 galaxies.

\section{Discussion} \label{sec: discussion}

S0 galaxies are known to be in general red and non-star forming (e.g., \cite{MM09, Bait17, Mishra19}). The number of S0 galaxies has been found to increase in proportion to the decrease in number of spirals, as we advance from the SFMS to the green valley and to the quenched sequence (e.g., \cite{Bait17}). We find a statistically significant sample of actively star-forming S0s, and thus natural questions that arise are - what is the site, extent and cause of star-formation in these objects? What is the evolutionary history of such elusive objects? Are these star-forming S0s actually fading spirals that will eventually become quenched S0s (see \cite{Johnston14})? Or are they S0s that were quenched in the past but have been somehow rejuvenated in star-formation? Or are they objects with a distinct evolutionary history of their own?

Based on Fig. \ref{fig: sigma_sfr_profile_with_typical_errors} and \ref{fig: ssfr_profile_with_typical_errors} in Section \ref{sec: results}, we observe that star-formation in SF-S0s is centrally dominated, as opposed to disc dominated star-formation in SF-Sps. $\Sigma_{SFR}$ for SF-S0s falls of with distance relatively more rapidly than that in SF-Sps. The central regions of SF-S0 galaxies have star-formation comparable to, or more than that of SF-Sps. Whereas, in the outer regions, SF-S0s are relatively less star-forming as compared to SF-Sps.

Next, we do a comparison of some of the global properties of SF-Sps, SF-S0s and Q-S0s, in order to identify similarity and differences of our main sample of SF-S0s with the two control samples. This analysis is expected to provide clues regarding the evolutionary history of SF-S0s.

\subsection{Size, Mass and Morphology} \label{sec:size_mass_morph}

\begin{figure}
    \centering
    \includegraphics[width = 0.45\textwidth]{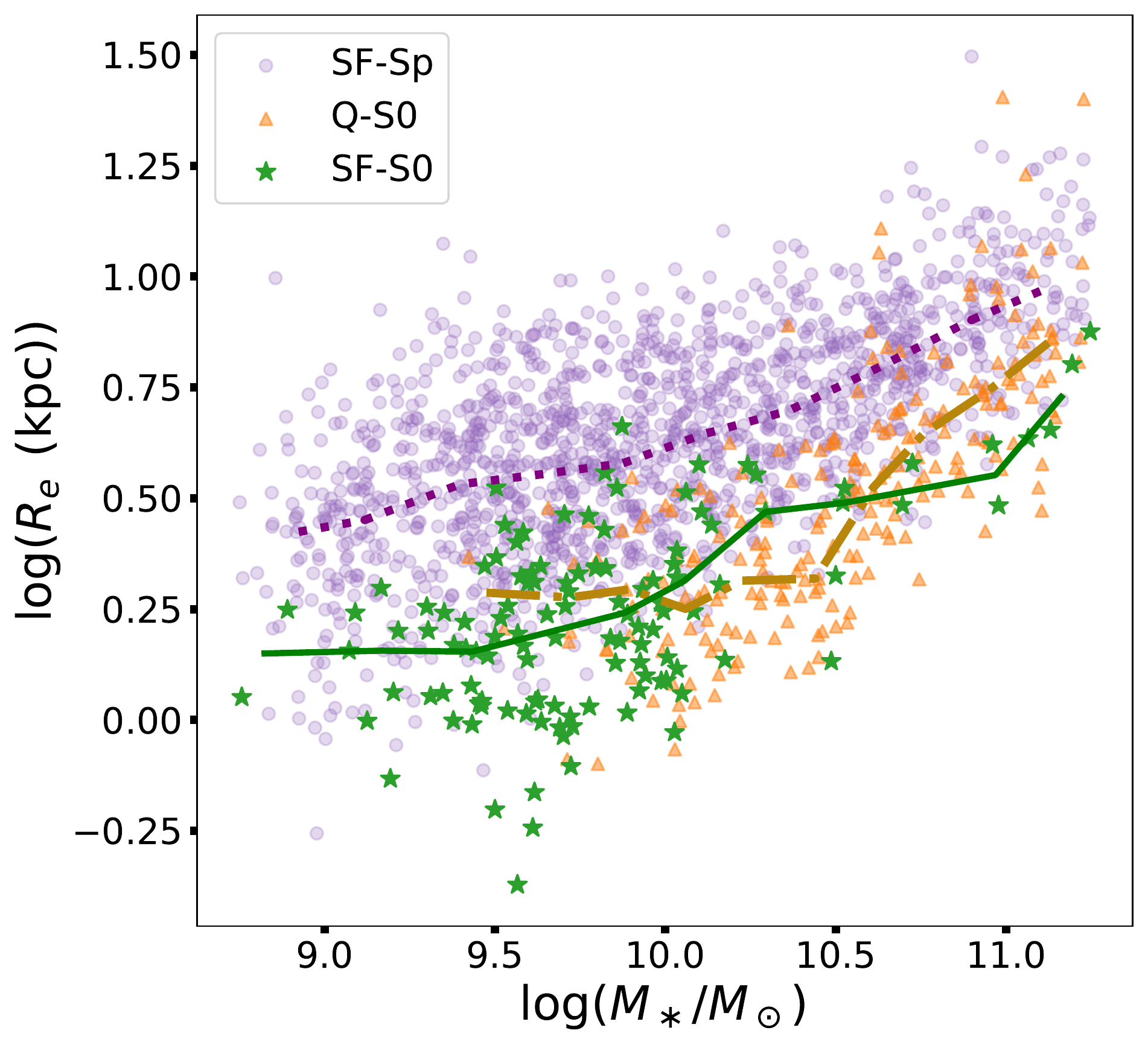}
    \caption{The size-mass relation for SF-Sps (purple circles), SF-S0s (green stars) and Q-S0s (orange triangles). Y-axis is the r-band half-light radius ($R_e$), and X-axis is the stellar mass. The solid lines represent the median trend for each sample}. The sample of SF-S0s seems to be more similar to Q-S0s in this parameter space, suggesting structural similarity between SF-S0s and Q-S0s.
    \label{fig:size_mass}
\end{figure}

In order to study the global structure of the SF-S0s, we construct the size-mass relation for the three samples.  The galaxy-size mass relation is an excellent tool for studying galaxy evolution \citep{Cappellari16} and has been used in previous works to compare galaxies of different colours and morphologies \citep{Wel14,Lange15,Mishra19}. The size-mass relation for our sample galaxies is shown in Fig. \ref{fig:size_mass}.

In Fig. \ref{fig:size_mass}, we show the scatter of the three samples in the size-mass parameter space. 
It is evident that SF-S0s are more similar to Q-S0s as compared to SF-Sps in this parameter space. This suggests that the global structure of SF-S0s is similar to Q-S0s on average, and the SF-S0s are structurally different from SF-Sps.

Interestingly, studies \citep{Mowla19,Kawinwanichakij19} have found a break in the size-mass relation at a stellar mass $\log{M_\ast/M_\odot} \sim 10.2$ (known as the \lq pivot mass\rq) for galaxies within $0.1<z<0.5$. The pivot mass is thought to be of physical significance as it is closer to the value of stellar mass threshold above which $50\%$ of the galaxies are quenched \citep{Mowla2019, Kawinwanichakij2021}, indicating different evolutionary histories for galaxies above and below pivot mass. In general, above this stellar mass threshold quenching becomes much more efficient \citep[e.g.,][]{Baldry06, Peng2010}. In our sample as well we find that the abundance of SF-S0s seems to decrease sharply beyond $\log{M_\ast/M_\odot} \sim 10.25$, as suggested by the stellar-mass histogram (Fig. \ref{fig: sm_hist_combined}). We speculate that such a correlation could be a manifestation of mass-driven quenching.

Bulge to total luminosity ratio ($B/T$) is a quantitative proxy for morphology. Thus, a larger $B/T$ indicates higher probability of a galaxy to be quenched. With this motivation, we compare $B/T$ values of our three samples, in order to obtain clues on whether the SF-S0s were a quenched population in the past. We take $B/T$ values of our galaxies from PM15, which are derived using S\'{e}rsic $+$ exponential fits. 
For the low stellar mass bin, we find the weighted median value of $B/T$ for SF-Sps to be $0.28$, for SF-S0s to be $0.53$ and for Q-S0s to be $0.51$. A weighted Kolomogorov-Smirnov (KS) test between SF-S0s and SF-Sps yields test statistic to be $0.38$ and p-value to be $\sim 10^{-12}$. The same test between SF-S0s and Q-S0s yields test statistic to be $0.18$ and p-value to be $0.18$. This clearly indicates that the null hypothesis that SF-S0s and SF-Sps are drawn from the same distribution, is rejected at a very high significance. For the high stellar mass bin, we find the weighted median value of $B/T$ for SF-Sps to be $0.18$, for SF-S0s to be $0.37$ and for Q-S0s to be $0.48$. Weighted KS test between SF-S0s and SF-Sps yields test statistic to be $0.66$ and p-value to be $\sim 10^{-5}$. The same test between SF-S0s and Q-S0s yields test statistic to be $0.28$ and p-value to be $0.33$. Again, the KS test reveals that the SF-S0s are different from the SF-Sps.

These findings re-enforce the hypothesis that the SF-S0s are different from spirals, and are more similar to Q-S0s. In particular, the hypothesis that $B/T$ distribution of SF-S0s is similar to SF-Sps is rejected at high significance by the KS test for both low mass and high mass galaxies.

\subsection{Kinematic Properties} \label{sec:kine}
Next, we study kinematic properties of our three samples, since properties like stellar velocity dispersion can yield useful comparison points between SF-Sps, SF-S0s and Q-S0s. We also inspect the stellar and gas kinematic maps of SF-S0s, since disturbed maps might suggest merger activity which can bring in fresh gas and potentially trigger star-formation.

\cite{Bluck16} have inferred that the central stellar velocity dispersion ($\sigma$) is a parameter that correlates the most with quenching of galaxies. A higher value of $\sigma$ is a very good predictor of quenching. With this motivation, we compute the stellar $\sigma$ for the three samples by constructing a $1$ bulge $R_e$ aperture on the stellar velocity dispersion maps provided by Pipe3D, and take mean over the spaxels lying inside that aperture. We take bulge $R_e$ values for our galaxies from S\'{e}rsic $+$ exponential fits of PM15. 

We cannot derive any conclusion based on stellar $\sigma$ for low stellar mass SF-Sp, SF-S0 and Q-S0 galaxies, since their values are small, and are similar within the typical error of $\sim 10$ km s$^{-1}$. In the high stellar mass bin, SF-S0s have large stellar $\sigma$ (weighted median being $128$ km s$^{-1}$) value which is very close to Q-S0s (weighted median being $126$ km s$^{-1}$), as compared to SF-Sps (weighted median being $67$ km s$^{-1}$). Weighted KS test between SF-S0s and SF-Sps yields the test statistic to be $0.43$ and p-value to be $0.03$. The same test between SF-S0s and Q-S0s yields test statistic to be $0.36$ and p-value to be $0.15$. 

These findings re-enforce the hypothesis that the high mass SF-S0s are different from SF-Sps and are more similar to Q-S0s, and were also likely quenched in the past.

We inspect stellar velocity and $H_\alpha$ gas velocity maps of SF-S0s using SDSS-MARVIN\footnote{\url{https://www.sdss.org/dr15/manga/marvin/}} - a tool for visualizing MaNGA IFS data \citep{Marvin}. Based on this visual inspection, we divide the SF-S0s into $4$ categories based on their kinematic maps:
\begin{itemize}
    \item Regular: stellar and $H_\alpha$ velocity maps show clean rotation, and both maps are aligned. There are $48$ such objects in our sample
    \item Disturbed: either one, or both of the stellar and $H_\alpha$ velocity maps are disturbed. There are $56$ such objects in our sample
    \item Misaligned: stellar and $H_\alpha$ velocity maps show rotation, but the axis of rotation of stars and gas seems to be misaligned. There are $3$ such objects in our sample
    \item Counter-rotating: stellar and $H_\alpha$ velocity maps show rotation, but the axis of rotation of stars and gas seems anti-aligned. There are $11$ such objects in our sample
\end{itemize}

\begin{figure*}
    \centering
    \includegraphics[width = 0.45\textwidth]{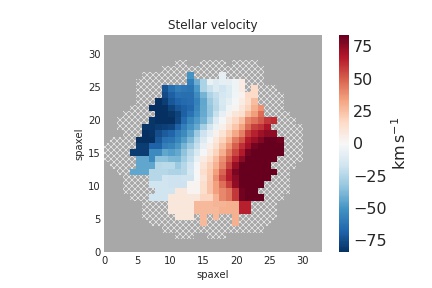}
    \includegraphics[width = 0.45\textwidth]{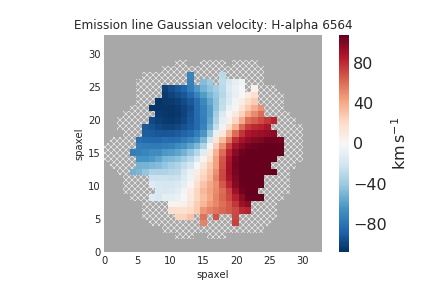}\\
    \includegraphics[width = 0.45\textwidth]{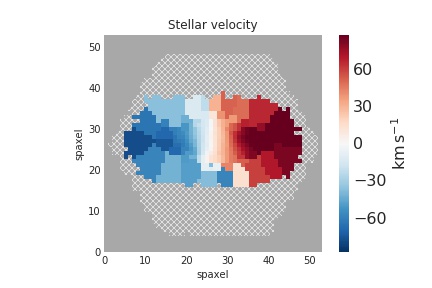}
    \includegraphics[width = 0.45\textwidth]{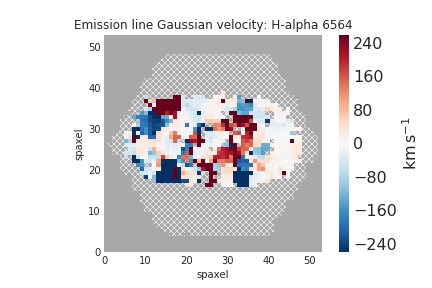}\\
    \includegraphics[width = 0.45\textwidth]{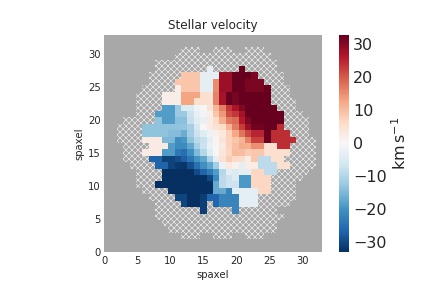}
    \includegraphics[width = 0.45\textwidth]{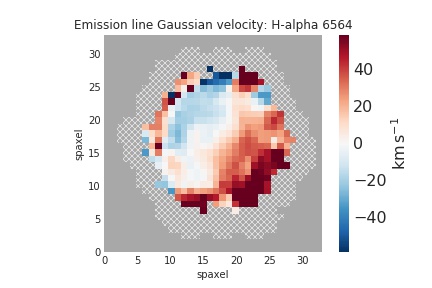}\\
    \includegraphics[width = 0.45\textwidth]{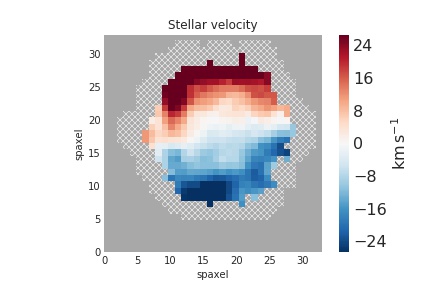}
    \includegraphics[width = 0.45\textwidth]{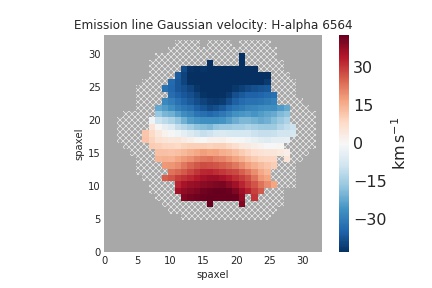}\\
    \caption{Example stellar velocity maps (left) and $H_\alpha$ velocity maps (right) of regular, disturbed, misaligned and counter-rotating galaxies. PLATEIFU (top to bottom): 8258-1902 ($12$ arcsec diameter), 8934-6101 ($22$ arcsec diameter), 8084-1902 ($12$ arcsec diameter ) and 9865-1901 ($12$ arcsec diameter). These maps have been produced using SDSS-MARVIN.}
    \label{fig:kinematics}
\end{figure*}

We would like to point out that these categories are based on just a visual inspection, and a counter-rotating object might turn out to be strongly mis-aligned, but not exactly anti-aligned. Also, only $3$ SF-S0s have inclination angle less than $15^{\circ}$, so only these $3$ objects can be said to be face-on. $1$ of them is in the regular category and $2$ are in disturbed. So, the contamination due to face-on objects in the disturbed category is very small. Example maps from each of the above $4$ categories are displayed in Fig. \ref{fig:kinematics}.

We have $14$ objects which are mis-aligned or counter-rotating. Gas accretion through recent mergers are the most likely cause for this \citep{Bendo00,Matteo07,Bassett17}. From our purposes, we define kinematically \lq unsettled\rq objects as those which are either disturbed, or misaligned, or counter-rotating. Overall, more than $50\%$ of our sample comprises of unsettled objects. These unsettled objects have likely undergone recent interactions or mergers that might have brought in fresh gas for star-formation. 

\begin{figure}
    \centering
    \includegraphics[width = 0.45\textwidth]{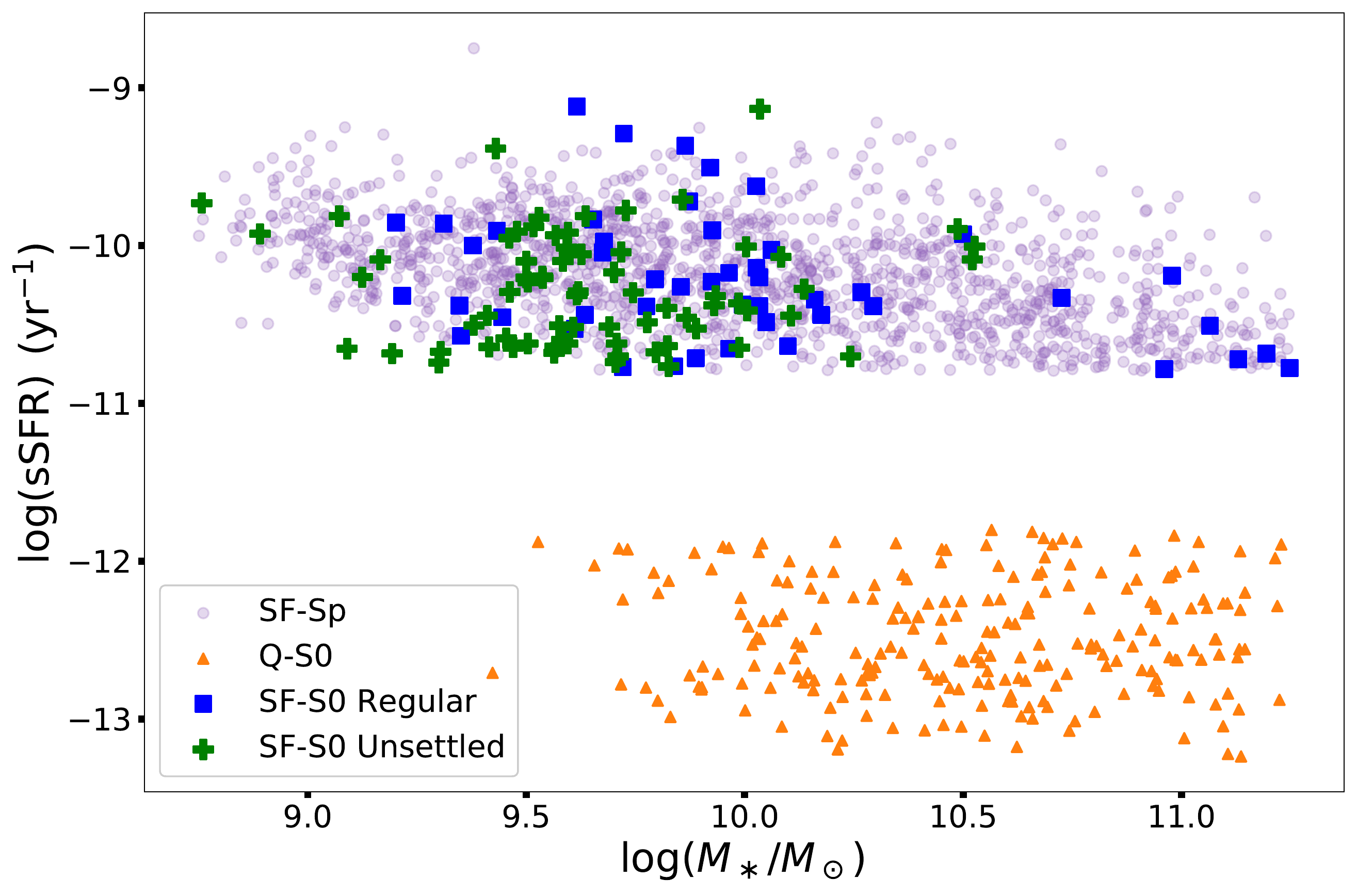}
    \caption{The Star Formation Main Sequence (SFMS) plot for the three samples. The control sample of star-forming spirals (SF-Sps) having $1468$ objects is depicted by light-purple circles, and the control sample of quenched S0s (Q-S0s) having $227$ objects is depicted by orange triangles. We further categorise the main sample of SF-S0s based on stellar and $H_\alpha$ kinematic maps, as explained in the text (Section \ref{sec:kine}). At the very low stellar mass end, most of the SF-S0 galaxies are kinematically unsettled. Whereas, at the high stellar mass end, most of the SF-S0 galaxies are kinematically regular.}
    \label{fig:SFMS_kine}
\end{figure}

Next, we try to identify whether correlations exist between the kinematic category, SFR and stellar mass of the SF-S0 galaxies. For this, we scatter the objects on the SFMS, and use different markers for different kinematic categories (see Fig. \ref{fig:SFMS_kine}). At the very low stellar mass end, most of the SF-S0 galaxies are unsettled. Whereas at the high stellar mass end, most of the SF-S0 galaxies are regular. We do not completely understand this effect, but speculate that it is in general difficult to kinematically perturb high stellar mass galaxies due to their already large rotational inertia, and hence most of them appear regular.
\\
\\
Evidences presented so far in this study suggest that the SF-S0s are structurally and morphologically similar to Q-S0s and are different from SF-Sps. This leads us to conclude that the SF-S0s are unlikely to be fading spirals. The SF-S0s were likely a quenched population in the past, and their star-formation has been rejuvenated. We find more than $50\%$ of the SF-S0 sample to be kinematically unsettled, thus minor mergers seem to be the most likely cause for rejuvenation. \cite{Kaviraj17} compared $NUV - r$ colours of intermediate mass blue early type galaxies at low redshift to synthetic photometry obtained from minor merger simulations. They conclude that minor mergers have a strong possibility for being responsible for recent star-formation in nearby blue early types, and point out that gas-cooling events in red early types are an unlikely cause due to virial shock heating in halos \citep{Dekel06}. 

Given the clear stellar mass difference between SF-S0s and Q-S0s (Fig. \ref{fig: sm_hist_combined}), we speculate that mass-driven quenching \citep{Peng2010} might also have a role to play in the evolutionary history as well as the subsequent evolution of SF-S0 galaxies. The SF-S0 galaxies might gain enough stellar mass as a result of active star-formation, and move to the quenched sequence due to mass-driven quenching. This might also explain why we find so few SF-S0s at higher stellar masses. However, if the duration of the rejuvenated star-formation in SF-S0s is short, they might become quenched without significant stellar mass buildup. 

The high stellar mass Q-S0s are a likely progenitor for the high stellar mass SF-S0s. The progenitors of the low stellar mass SF-S0s, which are likely quenched S0s at low stellar masses, are mostly not a part of our sample plausibly due to selection effects outlined in Section \ref{sec: global props}.

Larger samples which are preferably volume-limited and span a range of redshift will be needed to fully understand this elusive class of star-forming S0s. IFS data will be very helpful for such studies, since it will allow comparing properties of star-forming S0s with other classes of galaxies such as spirals and quenched S0s on local scales. But we caution that a key aspect in such works would be accurate identification of morphology. Spiral contamination will be significant in shallow imaging. The final data release of the SDSS-MaNGA survey might increase the sample size of SF-S0s, allowing better statistics. Moreover, cold gas observations of the SF-S0s will allow studying the efficiency and time-scales of star-formation in these objects, and hence constrain their subsequent evolution.

After our analysis was completed ,and when this paper was in the final stages of preparation, another study focusing on star-forming S0s in the SDSS-MaNGA survey \citep{Xu21} became available. Our independent work is significantly different as compared to \cite{Xu21} from the aspects of objectives, analysis methodology, sample selection and sample size. However, it is worth noting that \cite{Xu21} have also invoked some kind of a minor-merger driven rejuvenation phenomenon to explain the existence of star-forming S0s.

\section{Conclusion} \label{sec: conclusion}
In this work, we studied a sample of $120$ star-forming S0 (SF-S0) galaxies from the Integral Field Spectroscopic (IFS) SDSS-MaNGA survey. S0 galaxies are generally expected to be quenched and hence existence of star-forming S0s raises pointed questions with regards to their star-formation properties and evolutionary history. SF-S0 galaxies might be spirals that are fading into the quenched sequence, or they can be quenched S0s in which star-formation has somehow been rejuvenated. They might also be objects having a distinct evolutionary history of their own. We studied these enigmatic SF-S0 objects both on global as well as resolved scales, in order to uncover the site, extent and cause of star-formation. We constructed control samples of star-forming spirals (SF-Sps) and quenched S0s (Q-S0s) and compared properties of SF-S0s and the control sample objects with the goal of obtaining clues to the former's evolutionary history.

We selected SF-Sps, SF-S0s and Q-S0s based on specific Star Formation Rate (sSFR) and utilised the SDSS-MaNGA DR15 Deep Learning Morphology Catalogue (DL15) for morphological information. Since DL15 was based on shallow SDSS DR15 imaging, the sample of S0s might have contamination from spirals. Therefore, we carried out a strict visual inspection of SF-S0s and Q-S0s using deeper images, in order to identify and remove spiral contaminants. Finally, we obtained a curated and reliable sample of $120$ SF-S0s. All the galaxies we used have redshift in the range $0.01 < z < 0.075$. The samples were made volume limited by applying volume weights from the MaNGA Targeting Catalog, making our analysis more rigorous statistically.

Most of SF-S0s were found to have stellar mass $\log(M_\ast/M_\odot)\leq 10.25$ (see Fig. \ref{fig: sm_hist_combined}). Thus, we further divided the samples of SF-S0s, SF-Sps and Q-S0s into two stellar mass bins, and studied objects with $\log(M_\ast/M_\odot)\leq 10.25$ and $\log(M_\ast/M_\odot) > 10.25$ separately. 

We used publicly available Pipe3D data products for resolved maps of stellar population and emission line properties of MaNGA galaxies, and constructed radial profiles of $\Sigma_{SFR}$ (SFR surface density) and sSFR. We also analysed global properties like the size-mass relation, bulge to total light ($B/T$) ratio and stellar velocity dispersion of our three samples and inspected stellar and gas kinematic maps of the SF-S0s. Our main results may be summarised as follows 

\begin{itemize}
  \item $\Sigma_{SFR}$ and sSFR radial profiles (see Fig. \ref{fig: sigma_sfr_profile_with_typical_errors} and Fig. \ref{fig: ssfr_profile_with_typical_errors}) suggest that SF-S0s have centrally dominated star-formation, as opposed to disc-dominated star-formation in SF-Sps. Star-formation in the central regions of SF-S0s is comparable to or more than that of SF-Sps, but is relatively less in the outer regions as compared to SF-Sps
  \item The size-mass relation (see Fig. \ref{fig:size_mass}) suggests that SF-S0s are structurally more similar to Q-S0s and are different from SF-Sps. This leads to us to conclude that SF-S0s are unlikely to be fading spirals and were likely a quenched population in the past in which star-formation has been rejuvenated
  \item We find that SF-S0s typically have a high $B/T$ ratio which is similar to Q-S0s, as compared to SF-Sps. Moreover, the high stellar mass SF-S0s have high stellar velocity dispersion, which is very similar to Q-S0s, as compared to SF-Sps. $B/T$ and stellar velocity dispersion being high indicates quenching in galaxies and therefore these findings re-enforce the hypothesis that the SF-S0s were a quenched population in the past, in which star-formation has been rejuvenated. We performed the Kolmogorov -Smirnov statistical test to further quantify our findings, and the resulting test statistic and p-values are consistent with our hypothesis.
  \item We find that more than $50\%$ of the SF-S0s are kinematically unsettled, which is suggestive of recent interactions and mergers. Thus, we conclude that minor mergers were likely responsible for rejuvenating the star-formation in SF-S0s, by bringing in fresh gas
\end{itemize}

Most of the SF-S0s have low stellar masses, whereas most of the Q-S0s (see Fig. \ref{fig: sm_hist_combined}) have high stellar masses. This suggests mass-driven quenching to also play a role in the evolutionary history and subsequent evolution of SF-S0s. Our sample contains very few quenched S0s at low stellar masses possibly due to selection effects. Thus, we cannot study a representative population of the progenitors of low stellar mass SF-S0s with our sample.

Further in depth studies across stellar mass and morphology are required, particularly based on IFS data, in order to fully understand the rare and interesting class of SF-S0 galaxies. Atomic and molecular gas observations of the SF-S0s will be very helpful in constraining their evolution, since it might provide insights regarding time-scales and efficiency of star-formation in these objects.

\section*{Acknowledgements}
We are grateful to the anonymous referee for their thoughtful and constructive review of the paper and their suggestions which have considerably improved the paper. HR and KK would like to thank the National Initiative for Undergraduate Sciences (NIUS) program conducted by the Homi Bhabha Centre For Science Education, Mumbai for providing the opportunity and resources to work on this project and acknowledge the support of the Department of Atomic Energy, Govt. Of India, under Project Identification No. RTI4001. HR and KK also acknowledge the National Centre for Radio Astrophysics - Tata Institute of Fundamental Research, Pune for hosting them in December 2019.  PKM, YW and OB acknowledge the support of the Department of Atomic Energy, Government of India, under project no. 12-R\&D-TFR5.02-0700. This research made use of Astropy,\footnote{\url{http://www.astropy.org}} a community-developed core Python package for Astronomy \citep{astropy:2013, astropy:2018}; Photutils, an Astropy package for detection and photometry of astronomical sources \citep{photutils1.0.1}.

Funding for the Sloan Digital Sky
Survey IV has been provided by the
Alfred P. Sloan Foundation, the U.S.
Department of Energy Office of
Science, and the Participating
Institutions. SDSS-IV acknowledges support and
resources from the Center for High
Performance Computing  at the
University of Utah. The SDSS
website is www.sdss.org.
SDSS-IV is managed by the
Astrophysical Research Consortium
for the Participating Institutions
of the SDSS Collaboration including
the Brazilian Participation Group,
the Carnegie Institution for Science,
Carnegie Mellon University, Center for
Astrophysics | Harvard \&
Smithsonian, the Chilean Participation
Group, the French Participation Group,
Instituto de Astrof\'isica de
Canarias, The Johns Hopkins
University, Kavli Institute for the
Physics and Mathematics of the
Universe (IPMU) / University of
Tokyo, the Korean Participation Group,
Lawrence Berkeley National Laboratory,
Leibniz Institut f\"ur Astrophysik
Potsdam (AIP),  Max-Planck-Institut
f\"ur Astronomie (MPIA Heidelberg),
Max-Planck-Institut f\"ur
Astrophysik (MPA Garching),
Max-Planck-Institut f\"ur
Extraterrestrische Physik (MPE),
National Astronomical Observatories of
China, New Mexico State University,
New York University, University of
Notre Dame, Observat\'ario
Nacional / MCTI, The Ohio State
University, Pennsylvania State
University, Shanghai
Astronomical Observatory, United
Kingdom Participation Group,
Universidad Nacional Aut\'onoma
de M\'exico, University of Arizona,
University of Colorado Boulder,
University of Oxford, University of
Portsmouth, University of Utah,
University of Virginia, University
of Washington, University of
Wisconsin, Vanderbilt University,
and Yale University.

The Pan-STARRS1 Surveys (PS1) and the PS1 public science archive have been made possible through contributions by the Institute for Astronomy, the University of Hawaii, the Pan-STARRS Project Office, the Max-Planck Society and its participating institutes, the Max Planck Institute for Astronomy, Heidelberg and the Max Planck Institute for Extraterrestrial Physics, Garching, The Johns Hopkins University, Durham University, the University of Edinburgh, the Queen's University Belfast, the Harvard-Smithsonian Center for Astrophysics, the Las Cumbres Observatory Global Telescope Network Incorporated, the National Central University of Taiwan, the Space Telescope Science Institute, the National Aeronautics and Space Administration under Grant No. NNX08AR22G issued through the Planetary Science Division of the NASA Science Mission Directorate, the National Science Foundation Grant No. AST-1238877, the University of Maryland, Eotvos Lorand University (ELTE), the Los Alamos National Laboratory, and the Gordon and Betty Moore Foundation.

The Legacy Surveys consist of three individual and complementary projects: the Dark Energy Camera Legacy Survey (DECaLS; Proposal ID \#2014B-0404; PIs: David Schlegel and Arjun Dey), the Beijing-Arizona Sky Survey (BASS; NOAO Prop. ID \#2015A-0801; PIs: Zhou Xu and Xiaohui Fan), and the Mayall z-band Legacy Survey (MzLS; Prop. ID \#2016A-0453; PI: Arjun Dey). DECaLS, BASS and MzLS together include data obtained, respectively, at the Blanco telescope, Cerro Tololo Inter-American Observatory, NSF’s NOIRLab; the Bok telescope, Steward Observatory, University of Arizona; and the Mayall telescope, Kitt Peak National Observatory, NOIRLab. The Legacy Surveys project is honored to be permitted to conduct astronomical research on Iolkam Du’ag (Kitt Peak), a mountain with particular significance to the Tohono O’odham Nation.

NOIRLab is operated by the Association of Universities for Research in Astronomy (AURA) under a cooperative agreement with the National Science Foundation.

This project used data obtained with the Dark Energy Camera (DECam), which was constructed by the Dark Energy Survey (DES) collaboration. Funding for the DES Projects has been provided by the U.S. Department of Energy, the U.S. National Science Foundation, the Ministry of Science and Education of Spain, the Science and Technology Facilities Council of the United Kingdom, the Higher Education Funding Council for England, the National Center for Supercomputing Applications at the University of Illinois at Urbana-Champaign, the Kavli Institute of Cosmological Physics at the University of Chicago, Center for Cosmology and Astro-Particle Physics at the Ohio State University, the Mitchell Institute for Fundamental Physics and Astronomy at Texas A\&M University, Financiadora de Estudos e Projetos, Fundacao Carlos Chagas Filho de Amparo, Financiadora de Estudos e Projetos, Fundacao Carlos Chagas Filho de Amparo a Pesquisa do Estado do Rio de Janeiro, Conselho Nacional de Desenvolvimento Cientifico e Tecnologico and the Ministerio da Ciencia, Tecnologia e Inovacao, the Deutsche Forschungsgemeinschaft and the Collaborating Institutions in the Dark Energy Survey. The Collaborating Institutions are Argonne National Laboratory, the University of California at Santa Cruz, the University of Cambridge, Centro de Investigaciones Energeticas, Medioambientales y Tecnologicas-Madrid, the University of Chicago, University College London, the DES-Brazil Consortium, the University of Edinburgh, the Eidgenossische Technische Hochschule (ETH) Zurich, Fermi National Accelerator Laboratory, the University of Illinois at Urbana-Champaign, the Institut de Ciencies de l’Espai (IEEC/CSIC), the Institut de Fisica d’Altes Energies, Lawrence Berkeley National Laboratory, the Ludwig Maximilians Universitat Munchen and the associated Excellence Cluster Universe, the University of Michigan, NSF’s NOIRLab, the University of Nottingham, the Ohio State University, the University of Pennsylvania, the University of Portsmouth, SLAC National Accelerator Laboratory, Stanford University, the University of Sussex, and Texas A\&M University.

BASS is a key project of the Telescope Access Program (TAP), which has been funded by the National Astronomical Observatories of China, the Chinese Academy of Sciences (the Strategic Priority Research Program “The Emergence of Cosmological Structures” Grant \# XDB09000000), and the Special Fund for Astronomy from the Ministry of Finance. The BASS is also supported by the External Cooperation Program of Chinese Academy of Sciences (Grant \# 114A11KYSB20160057), and Chinese National Natural Science Foundation (Grant \# 11433005).
The Legacy Survey team makes use of data products from the Near-Earth Object Wide-field Infrared Survey Explorer (NEOWISE), which is a project of the Jet Propulsion Laboratory/California Institute of Technology. NEOWISE is funded by the National Aeronautics and Space Administration. The Legacy Surveys imaging of the DESI footprint is supported by the Director, Office of Science, Office of High Energy Physics of the U.S. Department of Energy under Contract No. DE-AC02-05CH1123, by the National Energy Research Scientific Computing Center, a DOE Office of Science User Facility under the same contract; and by the U.S. National Science Foundation, Division of Astronomical Sciences under Contract No. AST-0950945 to NOAO.

The Hyper Suprime-Cam (HSC) collaboration includes the astronomical communities of Japan and Taiwan, and Princeton University. The HSC instrumentation and software were developed by the National Astronomical Observatory of Japan (NAOJ), the Kavli Institute for the Physics and Mathematics of the Universe (Kavli IPMU), the University of Tokyo, the High Energy Accelerator Research Organization (KEK), the Academia Sinica Institute for Astronomy and Astrophysics in Taiwan (ASIAA), and Princeton University. Funding was contributed by the FIRST program from Japanese Cabinet Office, the Ministry of Education, Culture, Sports, Science and Technology (MEXT), the Japan Society for the Promotion of Science (JSPS), Japan Science and Technology Agency (JST), the Toray Science Foundation, NAOJ, Kavli IPMU, KEK, ASIAA, and Princeton University.

Based [in part] on data collected at the Subaru Telescope and retrieved from the HSC data archive system, which is operated by Subaru Telescope and Astronomy Data Center at National Astronomical Observatory of Japan.

\section*{Data availability}
All the data we use in this study, are available in the public domain. The catalogue of star-forming S0s (SF-S0s), star-forming spirals (SF-Sps) and quenched S0s (Q-S0s) that we construct are available as supplementary material in the online version of this paper.

\bibliographystyle{mnras}
\bibliography{references}
\bsp
\appendix

\section{Galaxy Images}
\begin{figure*}
    \centering
    \includegraphics[width = 1\textwidth]{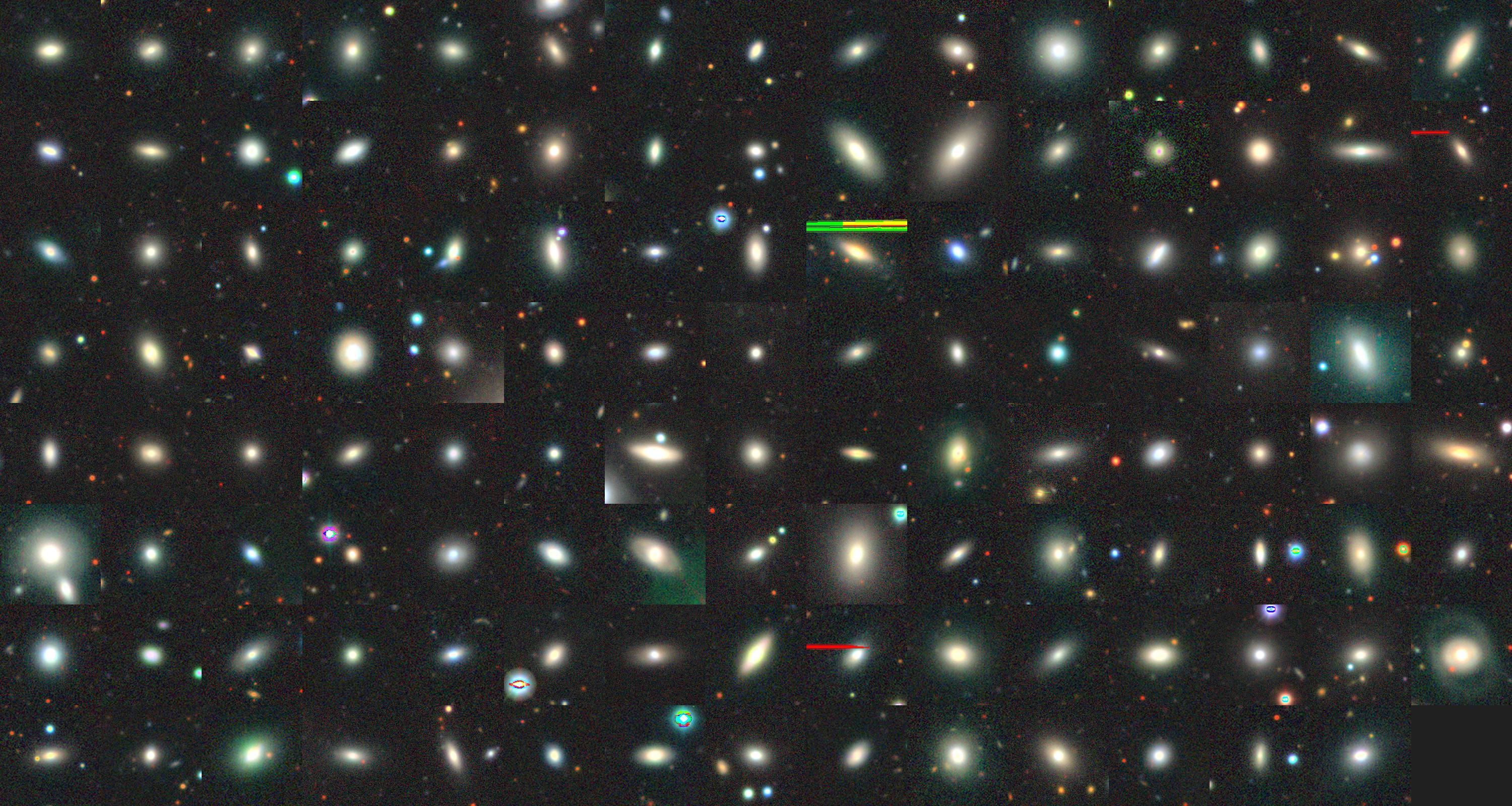}
    \caption{The DESI legacy imaging survey DR9 images of the SF-S0 galaxies. Each cutout is 45 arcsec by 45 arcsec.}
    \label{fig:fig_A1}
\end{figure*}
\begin{figure*}
    \centering
    \includegraphics[width = 0.5\textwidth]{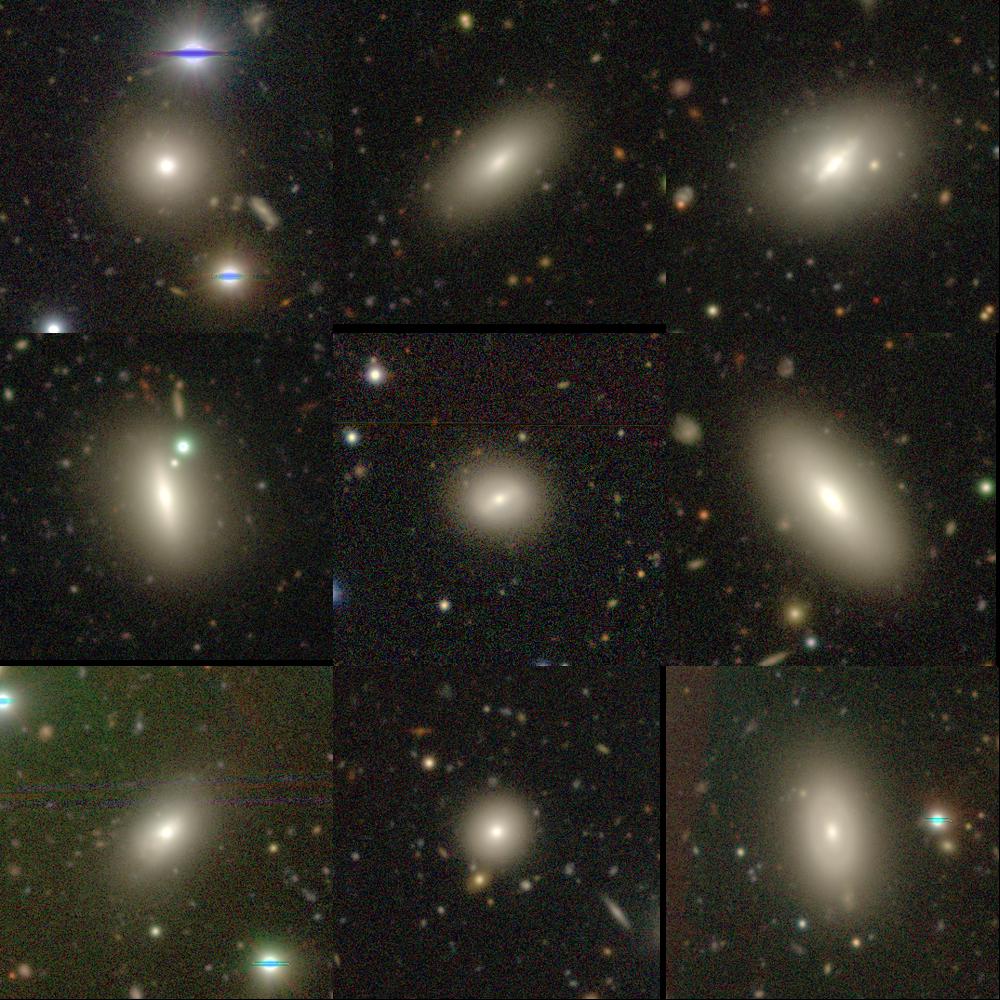}
    \caption{Nine of our SF-S0 galaxies which have  Hyper Suprime-Cam Subaru Strategic Program (HSC-SSP) PDR3 deep images. Dimension for each image is 40 arcsec $\times$ 40 arcsec.}
    \label{fig:fig_A2}
\end{figure*}
\begin{figure*}
    \centering
    \includegraphics[width = 0.3\textwidth]{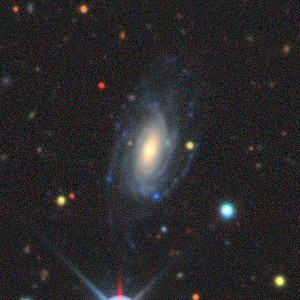}
    \includegraphics[width = 0.3\textwidth]{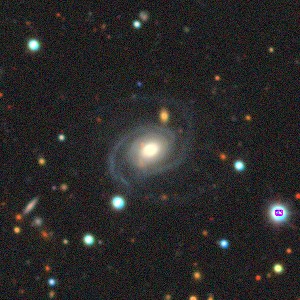}
    \caption{SDSS-DR14 images of two galaxies representing catastrophic failures in DL15 classification. These galaxies are clearly spirals, but are classified as S0s in DL15. PLATEIFUs (left and right): 8940-3703 and 9492-6102. Each image is 120 arcsec $\times$ 120 arcsec.}
    \label{fig:fig_A3}
\end{figure*}

\section{Data Tables}
\begin{table*}
    \centering
    \caption{Table of SF-S0 sample. \lq PLATEIFU\rq: plate-ifu of the MaNGA datacube, \lq RA\rq: Right Ascension, \lq DEC\rq: declination, \lq TTYPE\rq: Hubble TType value, \lq P\_S0\rq: measure of the probability of a galaxy being an S0 as opposed to being an elliptical, \lq z\rq: SDSS redshift, $\log(\frac{M_\ast}{M_\odot})$: logarithm of the stellar mass of the galaxy relative to solar, \lq log(SFR)\rq: logarithm of the star-formation rate of the galaxy (in $M_\odot$ yr$^{-1}$), $R_e$: half-light semi-major axis (arcsec), \lq B/A\rq: axis ratio (semi-minor/semi-major), \lq PA\rq: position angle (degrees), \lq Kine\_class\rq: kinematic category based on visual inspection of stellar and $H_\alpha$ velocity maps ($0$: regular, $1$: disturbed, $2$: mis-aligned, $3$: counter-rotating, $-1$: bad velocity data). TTYPE and P\_S0 have been taken from the MaNGA DR15 deep learning morphology catalog. Stellar mass, SFR and redshift have been taken from the GALEX-SDSS-WISE LEGACY catalog. $R_e$, B/A and PA have been taken from the r-band S\'{e}rsic fits of the MaNGA PyMorph DR15 catalogue. Typical errors (median error across all galaxies): $\Delta\log(\frac{M_\ast}{M_\odot}) = 0.04$, $\Delta\log(SFR) = 0.12$, $\Delta R_e = 0.03$ arcsec, $\Delta B/A = 0.004$, $\Delta PA = 0.53$ degrees}. The table has been sorted in increasing order of stellar mass. Full table available online.
    \begin{tabular}{c|c|c|c|c|c|c|c|c|c|c|c}
    \hline
     PLATEIFU   & RA (J2000)    & DEC (J2000)     & TTYPE & P\_S0 & z      & $\log(\frac{M_\ast}{M_\odot})$ & $\log$(SFR) & R$_e$   & B/A  & PA     & Kine\_class \\
        & (degrees)       & (degrees)     &  &  &       &  & ($M_\odot$ yr$^{-1}$) & (arcsec)   &   & (degrees)     &  \\
     \hline
    8144-1902  & 114.4580 & 28.6529 & -1.93 & 0.92  & 0.0158 & 8.757  & -0.97 & 3.39 & 0.89 & 8.39   & 1          \\
    9870-3703  & 232.9413 & 44.1081 & -1.33 & 0.84  & 0.0184 & 8.891  & -1.03 & 4.59 & 0.50 & 2.20   & 1          \\
    9488-1902  & 127.0117 & 20.9967 & -0.57 & 0.93  & 0.0182 & 9.072  & -0.74 & 3.76 & 0.86 & 45.80  & 1          \\
    8486-3703  & 238.8814 & 47.6773 & -0.03 & 0.94  & 0.0183 & 9.090  & -1.56 & 4.55 & 0.66 & -5.88  & 1          \\
    8933-9102  & 194.6532 & 27.8495 & -1.61 & 0.90  & 0.0206 & 9.125  & -1.08 & 2.31 & 0.97 & -86.74 & 1          \\
    8933-3704  & 195.3305 & 27.8605 & -0.48 & 0.94  & 0.0274 & 9.166  & -0.92 & 3.49 & 0.92 & 15.60  & 1          \\
    9195-1901  & 27.5653  & 13.1408 & -1.59 & 0.73  & 0.0161 & 9.193  & -1.49 & 2.18 & 0.87 & 25.59  & 1          \\
    9049-3704  & 247.5307 & 23.9260 & -0.52 & 0.98  & 0.0149 & 9.202  & -0.65 & 3.68 & 0.51 & 15.01  & 0          \\
    9486-1901  & 122.1268 & 39.8931 & -0.25 & 0.84  & 0.0209 & 9.216  & -1.10 & 3.64 & 0.74 & 41.20  & 0          \\
    8935-1901  & 195.9428 & 27.9869 & -2.07 & 0.95  & 0.0206 & 9.300  & -1.44 & 4.18 & 0.46 & -13.58 & 1          \\
    \hline
    \end{tabular}
    \label{tab:SFS0_Table}
\end{table*}

\begin{table*}
    \centering
    \caption{Table of our Q-S0 sample. Table columns and other details are as in Table B1. Typical errors (median error across all galaxies): $\Delta\log(\frac{M_\ast}{M_\odot}) = 0.02$, $\Delta\log(SFR) = 0.61$, $\Delta R_e = 0.04$ arcsec, $\Delta B/A = 0.002$, $\Delta PA = 0.25$ degrees}. Full Table available online.
    \begin{tabular}{c|c|c|c|c|c|c|c|c|c|c}
    \hline
    PLATEIFU   & RA (J2000)       & DEC (J2000)     & TTYPE & P\_S0 & z      & $\log(\frac{M_\ast}{M_\odot})$ & $\log$(SFR)   & R$_e$   & B/A  & PA     \\
        & (degrees)  & (degrees)     &  &  &       &  & ($M_\odot$ yr$^{-1}$)  &  (arcsec) &  & (degrees)  \\
    \hline
    9505-9101  & 138.5828 & 29.5457 & -1.87 & 0.80  & 0.0223 & 9.423  & -3.29 & 5.02  & 0.91 & 31.93  \\
    8983-3702  & 204.5444 & 26.7432 & -1.09 & 0.88  & 0.0279 & 9.527  & -2.35 & 2.78  & 0.79 & 77.05  \\
    8153-12703 & 41.0683  & 0.3416  & -1.75 & 0.61  & 0.0221 & 9.656  & -2.37 & 6.52  & 0.76 & 68.31  \\
    8335-1902  & 218.1676 & 40.9542 & -2.38 & 0.62  & 0.0181 & 9.712  & -2.21 & 2.15  & 0.49 & -11.27 \\
    8933-12703 & 196.3021 & 28.1524 & -1.62 & 0.95  & 0.0217 & 9.717  & -3.06 & 3.32  & 0.52 & 28.12  \\
    8625-1902  & 258.8203 & 57.3294 & -1.85 & 0.94  & 0.0262 & 9.721  & -2.52 & 2.91  & 0.47 & 16.51  \\
    8612-3702  & 253.4706 & 39.8296 & -2.25 & 0.64  & 0.0339 & 9.732  & -2.19 & 3.23  & 0.72 & 2.55   \\
    7991-6102  & 261.0818 & 56.8763 & -0.16 & 0.93  & 0.0275 & 9.774  & -3.03 & 4.93  & 0.39 & 78.80  \\
    8931-3702  & 194.0610 & 27.5063 & -0.80 & 0.86  & 0.0248 & 9.792  & -2.28 & 4.34  & 0.49 & 76.40  \\
    9031-1901  & 240.6183 & 44.5414 & -2.11 & 0.81  & 0.0207 & 9.801  & -3.08 & 5.32  & 0.54 & -11.73 \\
    \hline
    \end{tabular}
    \label{tab:QS0_Table}
\end{table*}

\begin{table*}
    \centering
    \caption{Table of our SF-Sp sample. Table columns and other details are as in Table B1. Typical errors (median error across all galaxies): $\Delta\log(\frac{M_\ast}{M_\odot}) = 0.04$, $\Delta\log(SFR) = 0.08$, $\Delta R_e = 0.05$ arcsec, $\Delta B/A = 0.002$, $\Delta PA = 0.28$ degrees}. Full Table available online.
    \begin{tabular}{c|c|c|c|c|c|c|c|c|c|c|c}
    \hline
    PLATEIFU   & RA (J2000)       & DEC (J2000)    & TTYPE  & z      & $\log(\frac{M_\ast}{M_\odot})$ & $\log$(SFR)   & R$_e$   & B/A  & PA     \\
       & (degrees)       & (degrees)     &  &  &       &  ($M_\odot$ yr$^{-1}$) & (arcsec)   &   & (degrees)     &  \\
    \hline
    8716-12701  & 119.6338 & 52.5980 & 2.88    & 0.0190 & 8.751  & -1.19 & 7.78  & 0.58 & 36.93  \\
    8716-12702  & 122.6800 & 52.5220 & 5.33    & 0.0189 & 8.760  & -1.07 & 5.27  & 0.78 & 10.10  \\
    8942-12703  & 125.2134 & 28.4059 & 5.67    & 0.0154 & 8.802  & -1.27 & 6.62  & 0.35 & 64.61  \\
    8982-12702  & 202.6234 & 26.5145 & 7.00    & 0.0250 & 8.810  & -0.75 & 7.82  & 0.29 & -48.27 \\
    9488-6102   & 126.6688 & 20.9673 & 2.27    & 0.0154 & 8.829  & -1.06 & 6.00  & 0.53 & -50.68 \\
    8566-6103   & 113.8983 & 41.9636 & 3.85    & 0.0103 & 8.836  & -1.13 & 4.74  & 0.62 & -13.36 \\
    9485-9102   & 121.0805 & 37.7094 & 3.82    & 0.0188 & 8.845  & -1.03 & 8.26  & 0.68 & 58.55  \\
    8728-6104   & 59.5806  & -4.7305 & 3.29    & 0.0177 & 8.846  & -0.96 & 7.35  & 0.85 & 49.51  \\
    8711-3703   & 119.1835 & 52.9890 & 1.40    & 0.0178 & 8.848  & -1.64 & 4.31  & 0.60 & -0.27  \\
    8987-6101   & 137.2349 & 27.5138 & 5.28    & 0.0219 & 8.848  & -1.04 & 6.06  & 0.15 & -89.72 \\
    \hline
    \end{tabular}
    \label{tab:SFSp_Table}
\end{table*}

\label{lastpage}
\end{document}